\documentclass[aps,prc,superscriptaddress,twoside,twocolumn,nofootinbib,10pt,%
showpacs,floatfix]{revtex4-1}

\usepackage{amsmath,amssymb}
\usepackage{graphicx,bm}
\usepackage{slashed}
\usepackage{epstopdf}
\usepackage{ulem} 
\usepackage[usenames]{color}
\usepackage{subfigure}
\usepackage{accents}
\usepackage{float}

\newcommand{\Slash}[1]{{\ooalign{\hfil/\hfil\crcr$#1$}}}

\allowdisplaybreaks
\begin{document}

\title{Spectral function for $\bar{D}_{0}^{\ast}$ $(0^+)$ meson in isospin asymmetric nuclear matter with chiral partner structure}

\author{Daiki Suenaga}
\email{suenaga@hken.phys.nagoya-u.ac.jp}
\affiliation{Department of Physics,  Nagoya University, Nagoya, 464-8602, Japan}


\date{\today}

\newcommand\sect[1]{\emph{#1}---}
\begin{abstract}
We study a spectral function for $\bar{D}_0^*$ $(0^+)$ meson in symmetric and neutron-rich asymmetric nuclear matter from the viewpoint of the partial restoration of chiral symmetry. Nuclear matter is constructed by a parity doublet model with hidden local symmetry which can reproduce properties at normal nuclear matter density as well as those in the vacuum. $\bar{D}$ mesons are introduced by the chiral partner structure. Our results show that a mass of $\bar{D}$ ($0^-$) meson increases while a mass of $\bar{D}_0^*$ meson decreases at mean field level as we increase the baryon density, reflecting the partial restoration of chiral symmetry. In the spectral function for $\bar{D}_0^*$ meson, we find a threshold enhancement is remarkably enhanced which indicates this peak is an appropriate probe to observe the partial restoration of chiral symmetry in nuclear matter. We also find a resonance of $\bar{D}_0^*$ meson revives at higher density due to a narrowing of the phase space. In the spectral function in neutron-rich asymmetric nuclear matter, we observe the threshold enhancement in negatively-charged $\bar{D}_0^*$ meson channel stands at higher energy and its height is more enhanced compared to neutral $\bar{D}_0^*$ meson channel, due to the violation of isospin symmetry. On the other hand, the resonance of negatively-charged $\bar{D}_0^*$ meson is slightly suppressed compared to the neutral one.
\end{abstract}
\maketitle

\section{Introduction}

Investigating chiral symmetry at temperature and/or density is one of the most important subjects in QCD (Quantum Chromodynamics). Chiral symmetry is spontaneously broken in the vacuum driven by a nonzero value of vacuum expectation value (VEV) of $\bar{q}q$, and hadrons acquire their masses. This symmetry is, however, expected to be partially (incompletely) restored at temperature and density so that we can expect significant changes of hadron masses and other properties in such environments accordingly~\cite{Hatsuda:1994pi}.

At temperature, an {\it ab}-{\it initio} calculation called lattice QCD works and chiral symmetry and related topics are energetically studied. At density, on the other hand, it is not straightforward to make use of the lattice QCD due to the ``sign problem''. Then, understanding of change of chiral symmetry at density is still poor in comparison with at temperature~\cite{DeTar:2009ef}. Hence, studying on this subject is attracting attention both theoretically and experimentally~\cite{Hayano:2008vn}.

In this situation, an experiment to investigate chiral symmetry at density by means of a pionic nucleus spectroscopy was performed at GSI~\cite{Suzuki:2002ae}. The result yielded a reduction of the pion decay constant ($f_\pi$) at the normal nuclear matter density ($\rho_0$) as $f_\pi^{*2}(\rho_0)/f_\pi^2 \approx 0.64$ which shows a partial restoration of chiral symmetry at the normal nuclear matter density. To get corroboration of this result, it is necessary to consider another method independently.


In order to explore chiral symmetry at density, we propose anti-charmed mesons ($\sim\bar{c}q$) can be appropriate probes~\cite{Suenaga:2014dia,Suenaga:2014sga,Suenaga:2015daa,Harada:2016uca,Suenaga:2017deu}, since these mesons include two advantages as follows. Anti-charmed mesons possess large masses compared to a typical energy scale of QCD ($\Lambda_{\rm QCD}$) so that a global symmetry called the $SU(2)_S$ heavy quark spin symmetry emerges~\cite{HQSS}. Thanks to this symmetry, it is possible to treat anti-charmed mesons carrying different spins equally. 
The other advantage is that anti-charmed mesons contain only one light quark and they are belonging to a fundamental representation of $SU(2)_L\times SU(2)_R$ chiral group. This simple representation allows us to construct a concise Lagrangian interacting with light mesons. Note that we especially focus on anti-charmed mesons to avoid difficulties of annihilation processes in nuclear matter.

So far, studies on (anti-)charmed mesons at temperature and/or density have been done by several methods: chiral effective model~\cite{Mishra:2003se,Yasui:2009bz,Blaschke:2011yv,Yasui:2012rw,Sasaki:2014asa}, QCD sum rule~\cite{Hayashigaki:2000es,Hilger:2008jg,Azizi:2014bba,Wang:2015uya,Suzuki:2015est} , coupled channel approach~\cite{Lutz:2005vx,Tolos:2009nn,Gamermann:2010zz,JimenezTejero:2011fc,GarciaRecio:2011xt}, quark meson coupling model~\cite{Tsushima:1998ru}, and so on. Studies on isospin asymmetric medium have also be done~\cite{Mishra:2008cd,Kumar:2010gb,Kumar:2011ff,Chhabra:2016vhp}. A novel phenomenon associated with heavy-flavored quarks and hadrons called QCD Kondo effect was also advocated~\cite{Hattori:2015hka} (In terms of studies on heavy-flavor physics in medium, see Ref.~\cite{Hosaka:2016ypm} for a review and references therein).

In the present analysis, we particularly employ an idea of chiral partner structure for anti-charmed mesons~\cite{Nowak:1992um,Bardeen:1993ae}. In the context of chiral partner structure, a mass difference between positive-parity meson and negative-parity meson is generated by the spontaneous breakdown of chiral symmetry. Hence, the mass difference between them gets narrowed in which the chiral symmetry is partially restored. In this paper, we regard $\bar{D}_0^*$ ($2318$) ($0^+$) and $\bar{D}$ ($1869$) ($0^-$) as the partner to each other while $\bar{D}_1$ ($2427$) ($1^+$) and $\bar{D}^*$ ($2010$) ($1^-$) as the partner.

The main decay mode of $\bar{D}_0^*$ meson is $\bar{D}_0^*\to \bar{D}\pi$ such that we expect significant changes of a spectrum of $\bar{D}_0^*$ meson in nuclear matter since the masses of $\bar{D}_0^*$ and $\bar{D}$ mesons are related by the chiral partner structure. Therefore, we particularly explore a spectral function for $\bar{D}_0^*$ meson in nuclear matter paying attention to the partial restoration of chiral symmetry, in addition to the masses of $\bar{D}_0^*$ meson and $\bar{D}$ meson.

In Ref.~\cite{Suenaga:2017deu}, we studied a spectral function for $\bar{D}_0^*$ meson in isospin symmetric nuclear matter constructed by a linear sigma model. Although our results showed a clear signal of partial restoration of chiral symmetry in the spectral function for $\bar{D}_0^*$ meson, the linear sigma model is so simple that our analysis was restricted at lower density region ($\rho_B \lesssim 0.1$ fm$^{-3}$). In the present study, by utilizing a parity doublet model for describing nuclear matter properly, we are allowed to explore the anti-charmed mesons around and higher than the normal nuclear matter density respecting chiral symmetry in our calculations. Furthermore, we are able to study the spectral function for $\bar{D}_0^*$ meson in isospin asymmetric nuclear matter with respect to chiral symmetry. These studies provide useful information of partial restoration of chiral symmetry for the future experiment such as the PANDA experiment at FAIR. Moreover, They are expected to be necessary to understand the collective behaviors in low energy heavy ion collisions planed in the CBM experiment at FAIR, or the J-PARC-HI program at J-PARC, and so on.

As stated above, nuclear matter is constructed by a parity doublet model. Within this model, positive-parity nucleon ($N(939)$) and negative-parity nucleon ($N^*(1535)$) is related by chiral dynamics and the masses of them get degenerated when chiral symmetry is restored as the chiral partner structure~\cite{Detar:1988kn,Nemoto:1998um,Jido:1998av,Jido:2001nt,Jido:1999hd}. In the present study, in particular, we follow the procedure done in Refs.~\cite{Motohiro:2015taa,Suenaga:2017wbb}. In these references, properties of nuclear matter such as the saturation density, the binding energy of a nucleon, the incompressibility and the symmetry energy as well as the vacuum properties are successfully reproduced by extending the parity doublet model. In order to describe isospin asymmetric nuclear matter, $\rho$ meson are included by employing the technique of hidden local symmetry (HLS)~\cite{Harada:2003jx}.

This paper is organized as follows: In Sec.~\ref{sec:PDModel}, we introduce the parity doublet model and construct isospin symmetric and asymmetric nuclear matter. In Sec.~\ref{sec:HMET}, a Lagrangian for anti-charmed mesons in a relativistic form based on the heavy quark spin symmetry and the chiral partner structure is derived. $\rho$ meson and $\omega$ meson are also incorporated into the Lagrangian by HLS. In Sec.~\ref{sec:Results}, calculations and results are shown. In Sec.~\ref{sec:Conclusion}, we provide conclusions and discussions.

\section{Parity doublet model}
\label{sec:PDModel}

In this section, we provide a Lagrangian for the nucleon ($N(939)$) and $N^*(1535)$ by the parity doublet model~\cite{Detar:1988kn,Jido:2001nt} and the hidden local symmetry (HLS)~\cite{Harada:2003jx}, and construct (asymmetric) nuclear matter. Here, we follow the procedure done in Refs.~\cite{Motohiro:2015taa,Suenaga:2017wbb} to determine model parameters.

\subsection{Construction of Lagrangian}
\label{sec:PDLagrangian}
In this subsection, we derive a Lagrangian for the nucleon and  $N^*(1535)$ interacting with light-flavored mesons. Baryons are not elementally particles such that it is possible to introduce two types of nucleon fields $\psi_{1}$ (naive-type) and $\psi_2$ (mirror-type) which transform under the $SU(2)_L \times SU(2)_R$ chiral transformation as
\begin{eqnarray}
\psi_{1,l} \to g_L\psi_{1,l}  \ &,& \ \psi_{1,r} \to g_R \psi_{1,r} \nonumber\\
\psi_{2,l} \to g_R\psi_{1,l}  \ &,& \ \psi_{2,r} \to g_L \psi_{2,r}\ . \label{ChiralTrans}
\end{eqnarray}
$\psi_{1,l}$, $\psi_{1,r}$, $\psi_{2,l}$ and $\psi_{2,r}$ are defined by 
\begin{eqnarray}
\psi_{1(2),l} &=& \frac{1-\gamma_5}{2}\psi_{1(2)}  \ , \nonumber\\
\psi_{1(2),r} &=& \frac{1+\gamma_5}{2}\psi_{1(2)} \ ,
\end{eqnarray}
and $g_L$ ($g_R$) is an element of $SU(2)_L$ ($SU(2)_R$) chiral group. By utilizing the transformation laws in Eq.~(\ref{ChiralTrans}), the Lagrangian for $\psi_1$ and $\psi_2$ interacting with $\sigma$ meson and pion respecting (gauged-)$[SU(2)_L\times SU(2)_R]_{\rm global}$ chiral symmetry, parity and charge conjugation can be obtained as
\begin{eqnarray}
{\cal L}_N &=& \bar{\psi}_{1r}i\Slash{D}\psi_{1r}+\bar{\psi}_{1l}i\Slash{D}\psi_{1l}  \nonumber\\
&+&  \bar{\psi}_{2r}i\Slash{D}\psi_{2r}+\bar{\psi}_{2l}i\Slash{D}\psi_{2l} \nonumber\\
&-& m_0\left[\bar{\psi}_{1l}\psi_{2r}-\bar{\psi}_{1r}\psi_{2l}-\bar{\psi}_{2l}\psi_{1r}+\bar{\psi}_{2r}\psi_{1l}\right]  \nonumber\\
&-& g_1\left[\bar{\psi}_{1r}M^\dagger\psi_{1l}+\bar{\psi}_{1l}M\psi_{1r}\right] \nonumber\\
&-& g_2\left[\bar{\psi}_{2r}M\psi_{2l}+\bar{\psi}_{2l}M^\dagger\psi_{2r}\right] \ .
 \label{PD1}
\end{eqnarray}
In Eq.~(\ref{PD1}), $m_0$, $g_1$ and $g_2$ are parameters and the covariant derivatives $D_\mu\psi_{1,l}$, $D_\mu\psi_{1,r}$, $D_\mu\psi_{2,l}$ and $D_\mu\psi_{2,r}$ are
\begin{eqnarray}
D_\mu \psi_{1,l(2,r)} &=& (\partial_\mu-i{\cal L}_\mu)\psi_{1,l (2,r)} \ , \nonumber\\
 D_\mu \psi_{1,r(2,l)} &=& (\partial_\mu-i{\cal R}_\mu)\psi_{1,r (2,l)}\ ,
 \end{eqnarray}
where ${\cal L}_\mu$ and ${\cal R}_\mu$ are external fields and the transformation laws are given by
\begin{eqnarray}
{\cal L}_\mu &\to& g_L{\cal L}_\mu g_L^\dagger -i\partial_\mu g_Lg_L^\dagger\ ,\nonumber\\
{\cal R}_\mu &\to& g_R{\cal R}_\mu g_R^\dagger -i\partial_\mu g_Rg_R^\dagger\ .
\end{eqnarray}
$M$ is the chiral field including $\sigma$ meson and pion which transforms as
\begin{eqnarray}
M \to g_L M g_R^\dagger \ . \label{MTrans}
\end{eqnarray}
By utilizing the polar-decomposition form, we find
\begin{eqnarray}
M = \xi_L^\dagger \sigma\xi_R = \sigma \xi_L^\dagger \xi_R =\sigma U\ . \label{MDeco}
\end{eqnarray}
$\xi_L$ and $\xi_R$ are the nonlinear realization of pion fields, and in the context of HLS, these fields transform under the $[SU(2)_L\times SU(2)_R]_{\rm global} \times [U(2)_V]_{\rm local}$ transformation as
\begin{eqnarray}
\xi_L \to h_\omega h_\rho \xi_L g_L^\dagger\ ,\ \ \xi_R \to h_\omega h_\rho \xi_R g_R^\dagger\ ,
\end{eqnarray}
where $h_\rho$ and $h_\omega$ are elements of $[SU(2)_V]_{\rm local}$ and $[U(1)_V]_{\rm local}$, respectively. Note that in the unitary gauge in terms of HLS, $\xi_L$ and $\xi_R$ are reduced to $\xi_L^\dagger = \xi_R = {\rm e}^{i\pi^a\tau^a/2\bar{\sigma}}$ and $U={\rm e}^{i\pi^a\tau^a/\bar{\sigma}}$, where $\tau^a$ is the Pauli matrix ($a=1,2,3$) and $\bar{\sigma}$ is a vacuum expectation value (VEV) of $\sigma$ meson which is identical to  the pion decay constant $f_\pi$ in the vacuum.

Next, we shall construct a mesonic part of the Lagrangian. It is convenient to introduce the following 1-forms $\hat{\alpha}_{\perp\mu}$ and $\hat{\alpha}_{\parallel\mu}$:
\begin{eqnarray}
\hat{\alpha}_{\perp\mu} &=& \frac{1}{2i}(D_\mu\xi_R\xi_R^\dagger-D_\mu\xi_L\xi_L^\dagger) \ ,\nonumber\\
\hat{\alpha}_{\parallel\mu} &=& \frac{1}{2i}(D_\mu\xi_R\xi_R^\dagger+D_\mu\xi_L\xi_L^\dagger)\ ,  \label{MC1Form}
\end{eqnarray}
with
\begin{eqnarray}
D_\mu\xi_L &=& \partial_\mu\xi_L-iV_\mu\xi_L + i\xi_L{\cal L}_\mu \ ,\nonumber\\
D_\mu\xi_R &=& \partial_\mu\xi_R-iV_\mu\xi_R + i\xi_R{\cal R}_\mu \  , \nonumber\\  \label{CVXi} 
\end{eqnarray}
where we have introduced vector mesons field $V_\mu$ as
\begin{eqnarray}
V_\mu = \frac{g_\rho}{2}\rho_\mu + \frac{g_\omega}{2}\omega_\mu
\end{eqnarray}
($\rho_\mu=\rho_\mu^a\tau^a$). In the context of HLS, $\rho$ meson and $\omega$ meson are regarded as the gauge fields in terms of $[SU(2)_{V}]_{\rm local}$ and $[U(1)_V]_{\rm local}$ symmetry, respectively. Then the transformation laws for the vector mesons field $V_\mu$ is
\begin{eqnarray}
V_\mu \to h_\rho V_\mu h_\rho^\dagger -i\partial_\mu(h_\rho h_\omega) h_\rho^\dagger h_\omega^\dagger\ . \label{TransV}
\end{eqnarray}
Hence, by using Eqs.~(\ref{MC1Form}) -~(\ref{TransV}), the mesonic part of the Lagrangian can be found as
\begin{eqnarray}
{\cal L}_{\rm M} &=& \frac{1}{2}\partial_\mu\sigma\partial^\mu\sigma + \sigma^2{\rm Tr}[\hat{\alpha}_{\perp\mu}\hat{\alpha}_\perp^{\mu}] \nonumber\\
&+&  \frac{m_\rho^2}{g_\rho^2}{\rm Tr}[\hat{\alpha}_{\parallel\mu}\hat{\alpha}_{\parallel}^\mu]+\left(\frac{m_\omega^2}{2g_\omega^2}-\frac{m_\rho^2}{2g_\rho^2}\right){\rm Tr}[\hat{\alpha}_{\parallel\mu}]{\rm Tr}[\hat{\alpha}_\parallel^\mu] \nonumber\\
&-&\frac{1}{2g_\rho^2}{\rm Tr}[V_{\mu\nu}V^{\mu\nu}]-\left(\frac{1}{4g_\omega^2}-\frac{1}{4g_\rho^2}\right){\rm Tr}[V_{\mu\nu}]{\rm Tr}[V^{\mu\nu}] \nonumber\\
&+&\frac{\bar{\mu}^2}{2}\sigma^2-\frac{\lambda}{4}\sigma^4+\frac{\lambda_6}{6}\sigma^6 + \frac{\bar{m}\epsilon}{4}\sigma{\rm Tr}[U+U^\dagger]\ ,
\nonumber\\ \label{LMeson}
\end{eqnarray}
where we have defined the field strengths for vector mesons by
\begin{eqnarray}
V_{\mu\nu} = \partial_\mu V_\nu-\partial_\nu V_\mu-i[V_\mu,V_\nu]\ ,
\end{eqnarray}
and ``Tr'' stands for the trace for isospin index. In Eq.~(\ref{LMeson}), $g_\rho$, $g_\omega$, $\bar{\mu}$, $\lambda$, $\lambda_6$, $\bar{m}\epsilon$ are parameters which will be determined in Sec.~\ref{sec:NuclearMatter}. Note that six-point interaction of $\sigma$ meson is included in this expression. We should also note the last term in Eq.~(\ref{LMeson}) violating the chiral symmetry is added so as to reproduce the pion mass.

Furthermore, the interaction manners for $\psi_1$, $\psi_2$ and $\rho$, $\omega$ mesons respecting $({\rm gauged}$-$)[SU(2)_L\times SU(2)_R]_{\rm global} \times [U(2)_V]_{\rm local}$ symmetry are derived as
\begin{eqnarray}
{\cal L}_{NV} &=&a_{\rho}[\bar{\psi}_{1l}(\xi_L^\dagger\Slash{\hat{\alpha}}_{\parallel}\xi_L)\psi_{1l}+\bar{\psi}_{1r}(\xi_R^\dagger\Slash{\hat{\alpha}}_{\parallel}\xi_R)\psi_{1r}] \nonumber\\
&&\ \ \ \ \ + \bar{\psi}_{2l}(\xi_R^\dagger\Slash{\hat{\alpha}}_{\parallel}\xi_R)\psi_{2l}+\bar{\psi}_{2r}(\xi_L^\dagger\Slash{\hat{\alpha}}_{\parallel}\xi_L)\psi_{2r}] \nonumber\\
&+& \frac{a_{\omega}}{2}{\rm Tr}[\hat{\alpha}_{\parallel\mu}](\bar{\psi}_{1,l}\gamma^\mu\psi_{1,l}+\bar{\psi}_{1,r}\gamma^\mu\psi_{1,r} \nonumber\\
&&\ \ \ \ \ \ \  \ \ \ \ \ +\bar{\psi}_{2,l}\gamma^\mu\psi_{2,l}+\bar{\psi}_{2,r}\gamma^\mu\psi_{2,r})\ . \label{NVInt}
\end{eqnarray}
Under the spontaneous breakdown of chiral symmetry, $\sigma$ meson field acquires its VEV $\bar{\sigma}$. Hence, by changing the variable as $\sigma \to \bar{\sigma}+\sigma$ in Eqs.~(\ref{PD1}),~(\ref{LMeson}) and~(\ref{NVInt}), we can obtain the Lagrangians in the chiral broken phase.

The fermion fields $\psi_1$ and $\psi_2$ are not physical states since the mass matrix is not diagonalized.  The mass eigenstates $N_+$ and $N_-$ are given by mixing states of $\psi_1$ and $\psi_2$:
\begin{eqnarray}
\left(
\begin{array}{c}
N_+ \\
N_- \\
\end{array}
\right) = \left(
\begin{array}{cc}
{\rm cos}\, \theta & \gamma_5{\rm sin}\, \theta \\
-\gamma_5{\rm sin}\, \theta & {\rm cos}\, \theta \\
\end{array}
\right)\left(
\begin{array}{c}
\psi_1 \\
\psi_2\\
\end{array}
\right)\ , \label{Angle}
\end{eqnarray}
and corresponding mass eigenvalues are
\begin{eqnarray}
m_{N_+} &=& \frac{1}{2}\left(\sqrt{(g_1+g_2)^2\bar{\sigma}^2+4m_0^2}-(g_2-g_1)\bar{\sigma}\right) \ , \nonumber\\
m_{N_-} &=& \frac{1}{2}\left(\sqrt{(g_1+g_2)^2\bar{\sigma}^2+4m_0^2}+(g_2-g_1)\bar{\sigma}\right) \ .\nonumber\\
\end{eqnarray}
In Eq.~(\ref{Angle}), we have introduced a mixing angle $\theta$ which satisfies
\begin{eqnarray}
{\rm tan}\, 2\theta = \frac{2m_0}{(g_1+g_2)\bar{\sigma}}\ .
\end{eqnarray}
We note that $N_+$ carries positive parity while $N_-$ carries negative parity, so that we assign $N_+$ to the nucleon ($N(939)$) while $N_-$ to $N^*(1535)$ in our analysis.

\begin{table*}[htb]
  \begin{tabular}{cccccc||cccc} \hline\hline 
$\accentset{\circ}{m}_+$(MeV) & $\accentset{\circ}{m}_-$(MeV) & $\accentset{\circ}{m}_\omega$(MeV)&$\accentset{\circ}{m}_\rho$(MeV) & $\accentset{\circ}{m}_\pi$(MeV) & $f_\pi$(MeV) &$\rho_0$(fm$^{-3}$) & $E/A-\accentset{\circ}{m}_+$(MeV)& $K$ (MeV) & $S$ (MeV) \\\hline
939 & 1535   & 783 & 776& 140 & 93 &0.16 &-16 &240 & 31
\\ \hline \hline
  \end{tabular}
\caption{
Input parameters in our analysis. We take the saturation density, the binding energy of a nucleon, the incompressibility and the symmetry energy as inputs in addition to the nucleon mass, $N^*(1535)$ mass, $\omega$ meson mass, $\rho$ meson mass, pion mass and the pion decay constant.}
  \label{tab:Parameters}
\end{table*}
\begin{table*}[htb]
  \begin{tabular}{c||ccccccc} \hline\hline 
$m_0$ (MeV) &$g_1$& $g_2$ & $\bar{\mu}^2$ (MeV$^2$)& $\lambda$ & $\lambda_6$ (MeV$^{-2}$) & $g_{\omega NN}$ & $g_{\rho NN}$  \\\hline
$500$ &$8.96$ & $15.4$ & $1.92\times 10^5$ & $40.8$ & $19.1\times 10^{-3}$ & $11.4$ & $3.66$ \\
$700$ & $7.76$ & $14.2$ &$1.63\times10^5$ & $34.5$ & $1.56\times 10^{-3}$ & $7.31$ & $4.06$\\
 \hline \hline
  \end{tabular}
\caption{Determination of parameters for a given $m_0$. Here, we take $m_0=500$ MeV and $m_0=700$ MeV as examples.}
  \label{tab:CInputs}
\end{table*}
\subsection{Construction of nuclear matter}
\label{sec:NuclearMatter}

In this subsection, we construct (asymmetric) nuclear matter from the Lagrangians in Eqs.~(\ref{PD1}),~(\ref{LMeson}) and~(\ref{NVInt}). At a mean field approximation, by replacing $\omega_\mu \to\bar{\omega}\delta_{\mu0}$, $\rho_\mu^a \to \bar{\rho}\delta_{\mu0}\delta^{a3}$ and ${\cal L}_\mu = {\cal R}_\mu \to (\mu_B+\mu_I\tau^3)\delta_{0\mu}$, and defining $\bar{\rho}$ and $\bar{\omega}$ properly, we can find the thermodynamic potential per volume ($\Omega/V$) by performing the one-loop path integral in terms of $N_+$ and $N_-$ as
\begin{eqnarray}
&&\Omega/V \nonumber\\
&=& - \frac{1}{8\pi^2}\sum_{i=p,n} \theta(\mu_i^{*}-m_+)\nonumber\\
&& \left\{\frac{2}{3}E_{F,i}^{+}k_{F,i}^{+3}-E_{F,i}^{+}k_{F,i}^{+}m_+^2+m_+^4{\rm ln}\left(\frac{k_{F,i}^{+}+E_{F,i}^{+}}{m_+}\right)\right\} \nonumber\\
&-& (\rm negative\ parity ) \nonumber\\
&-& \left(\frac{\bar{\mu}^2}{2}\bar{\sigma}^2-\frac{\lambda}{4}\bar{\sigma}^4+\frac{\lambda_6}{6}\bar{\sigma}^6 + \bar{m}\epsilon\bar{\sigma} + \frac{m_\rho^2}{2}\bar{\rho}^2 + \frac{m_\omega^2}{2}\bar{\omega}^2\right)\ . \nonumber\\
\end{eqnarray}
In this equation, we have defined effective chemical potentials for the proton $p$ ($p^*$) and the neutron $n$ ($n^*$) as
\begin{eqnarray}
\mu_p^{*} &=&  \mu_B^*+\mu_I^* \ , \nonumber\\
\mu_n^{*} &=& \mu_B^*-\mu_I^* \ ,
\end{eqnarray}
respectively, and $\mu_B^*$ and $\mu_I^*$ are defined by 
\begin{eqnarray}
\mu_B^* &=& \mu_B-g_{\omega NN}\bar{\omega}\ , \nonumber\\
\mu_I^* &=& \mu_I-g_{\rho NN}\bar{\rho}\ ,
\end{eqnarray}
with $g_{\omega NN} = \frac{(a_\rho+a_\omega)g_\omega}{2}$ and $g_{\rho NN} = \frac{a_\rho g_\rho}{2}$. $E_{F,i}^{\pm}$ is defined by $E_{F,i}^{\pm} = \sqrt{k_{F,i}^{\pm2}+m_\pm^2}$, and $k^\pm_{F,i}$ is the Fermi momentum for the proton $p$ ($p^*$) or the neutron $n$ ($n^*$) defined by $\mu_i^*=\sqrt{k_{F,i}^{\pm2}+m_\pm^2}$ which is related to the density ($\rho_i^\pm$) by $\rho_i^\pm=\frac{1}{3\pi^2}k_{F,i}^{\pm3}$. The ground state of the system is determined by stationary conditions of $\Omega/V$ in terms of $\bar{\sigma}$, $\bar{\omega}$ and $\bar{\rho}$:
\begin{eqnarray}
\bar{\mu}^2\bar{\sigma}-\lambda\bar{\sigma}^3+\lambda_6\bar{\sigma}^5+\bar{m}\epsilon &=& \frac{\partial m_+}{\partial \bar{\sigma}}\rho_S^++\frac{\partial m_-}{\partial \bar{\sigma}}\rho_S^-\ , \label{StationarySigma}\\
 \bar{\omega} &=& \frac{g_{\omega NN}}{m_\omega^2}\rho_B \ ,\label{StationaryOmega}\\
 \bar{\rho} &=& \frac{g_{\rho NN}}{m_\rho^2}\rho_B \delta\ , \label{StationaryRho}
\end{eqnarray}
with
\begin{eqnarray}
\rho_S^{\pm} &=& 2\sum_{i=p,n}\int\frac{d^3k}{(2\pi)^3}\frac{m_{\pm}}{ \sqrt{|\vec{k}|^2+m_\pm^2}}\theta(k_{F,i}^{\pm }-|\vec{k}|) \ , \nonumber\\
\rho_B &=& \sum_{i=p,n}(\rho_i^++\rho_i^-) \ , \nonumber\\
\delta &=& \frac{\rho_p-\rho_n}{\rho_B}\ .
\end{eqnarray}
These equations indicate that the baryon number density $\rho_B$ and the asymmetry parameter $\delta$ are directly determined by $\bar{\omega}$ and $\bar{\rho}$, respectively.

In constructing (asymmetric) nuclear matter, we take the saturation density ($\rho_0$), the binding energy of a nucleon ($E/A-\accentset{\circ}{m}_+$) ($E$ is the total energy and $A$ is a mass number), the incompressibility $(K)$ and the symmetry energy ($S$) as inputs in addition to the nucleon mass ($\accentset{\circ}{m}_+$), $N^*(1535)$ mass ($\accentset{\circ}{m}_-$), $\omega$ meson mass ($\accentset{\circ}{m}_\omega$), $\rho$ meson mass ($\accentset{\circ}{m}_\rho$), pion mass ($\accentset{\circ}{m}_\pi$) and the pion decay constant ($f_\pi$)~\footnote{In this paper, we use the symbol ``$\accentset{\circ}{X}$'' to denote the vacuum value of $X$.}. These are summarized in Table.~\ref{tab:Parameters}.

\begin{figure}[htbp]
\centering
\includegraphics*[scale=0.4]{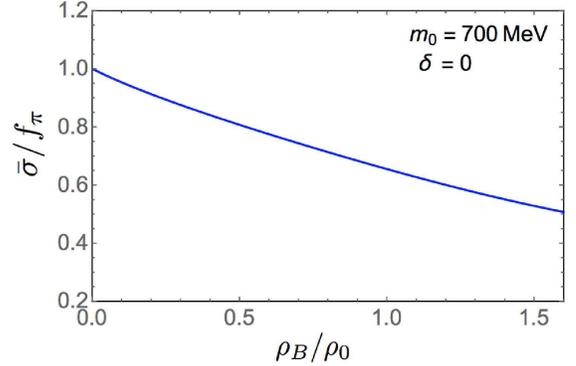}
\caption{A density dependence of $\bar{\sigma}$ with $m_0=700$ MeV and $\delta=0$. The value of $\bar{\sigma}$ decreases as the baryon number density increases which clearly shows the partial restoration of chiral symmetry.}
\label{fig:VEV700}
\end{figure}

We should note that all parameters except for $m_0$ is fixed, and only $m_0$ is remained as a free parameter in the present analysis. The resulting output parameters are listed in Table.~\ref{tab:CInputs}. In this table, we show the results for $m_0=500$ MeV and $m_0=700$ MeV as examples. A density dependence of $\bar{\sigma}$ with $m_0=700$ MeV and $\delta=0$ is plotted in Fig.~\ref{fig:VEV700}. As we can see, the value of $\bar{\sigma}$ decreases as the baryon number density increases which clearly shows the partial restoration of chiral symmetry in nuclear matter. 

Next, we shall show the way to treat fluctuations of $\sigma$ meson and pion in (asymmetric) nuclear matter. In the following, fluctuations of $\omega$ meson and $\rho$ meson will not be included since the masses of them are relatively large compared to $\sigma$ meson and pion. The true quantum field of $\sigma$ meson is defined by its fluctuation around the ground state determined by Eqs.~(\ref{StationarySigma}) -~(\ref{StationaryRho}). Then, by changing the variable as $\sigma \to \bar{\sigma}+{\sigma}$ , and performing a one-loop path integral in the original Lagrangian, the effective action for $\sigma$ meson and pion in the system is obtained:
\begin{eqnarray}
\Gamma &=& \Gamma_0[\bar{\sigma},\bar{\omega},\bar{\rho}] +  \Gamma_{\rm fluctuation}[\bar{\sigma},\bar{\omega},\bar{\rho}; {\sigma},{\pi}] \ ,
\end{eqnarray}
where $\Gamma_0[\bar{\sigma},\bar{\omega},\bar{\rho}]$ is the effective action at mean field level, and $\Gamma_{\rm fluctuation}[\bar{\sigma},\bar{\omega},\bar{\rho}; {\sigma},{\pi}]$ includes the fluctuations as well as the mean fields. Note that stationary condition of $\Gamma_0[\bar{\sigma},\bar{\omega},\bar{\rho}]$ with respect to $\bar{\sigma}$, $\bar{\omega}$ and $\bar{\rho}$ coincide with Eqs.~(\ref{StationarySigma}) -~(\ref{StationaryRho}).
\begin{figure*}[htbp]
\centering
\includegraphics*[scale=0.42]{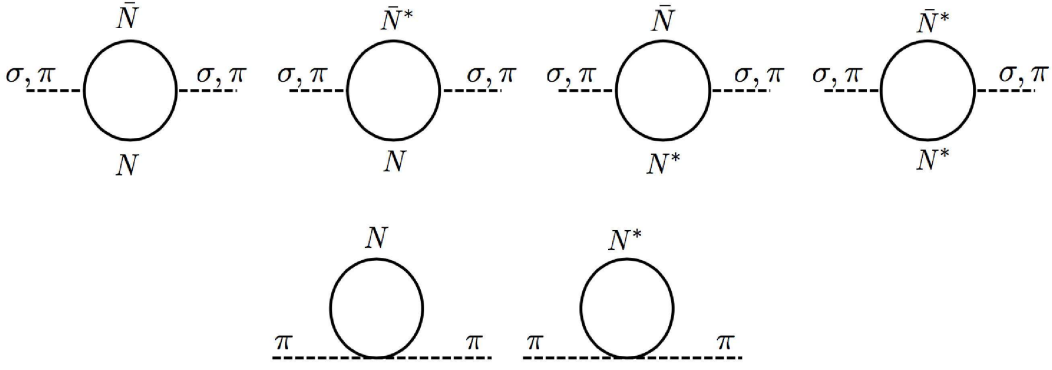}
\caption{A set of self-energies $\tilde{\Sigma}_{\sigma} (q_0,\vec{q})$, $\tilde{\Sigma}_{\pi^0} (q_0,\vec{q})$, $\tilde{\Sigma}_{\pi^+}(q_0,\vec{q})$ and $\tilde{\Sigma}_{\pi^-}(q_0,\vec{q})$.}
\label{fig:NNstLoops}
\end{figure*}

For later use, we need to derive two-point functions for $\sigma$ meson and pion. They are obtained by taking second functional derivatives with respect to $\sigma$ meson and pion:
\begin{eqnarray}
\tilde{G}_{\sigma}(q_0,\vec{q}) &=& i\left(\int d^4x\, {\rm e}^{iq\cdot x}\frac{\delta}{\delta\sigma(x)}\frac{\delta}{\delta\sigma(0)}{\Gamma}\right)^{-1} \nonumber\\
&\equiv&\frac{i}{q^2-m_{\sigma}^2-i\tilde{\Sigma}_{\sigma}(q_0,\vec{q})} \ ,  \label{TwoSigma} \\
\tilde{G}_{\pi^0}(q_0,\vec{q}) &=& i\left(\int d^4x\, {\rm e}^{iq\cdot x}\frac{\delta}{\delta\pi^0(x)}\frac{\delta}{\delta\pi^0(0)}{\Gamma}\right)^{-1} \nonumber\\
&\equiv&\frac{i}{q^2-m_{\pi^0}^2-i\tilde{\Sigma}_{\pi^0}(q_0,\vec{q})} \ ,  \label{TwoPi0} \\
\tilde{G}_{\pi^\pm}(q_0,\vec{q}) &=& i\left(\int d^4x\, {\rm e}^{iq\cdot x}\frac{\delta}{\delta\pi^\pm(x)}\frac{\delta}{\delta\pi^\mp(0)}{\Gamma}\right)^{-1}\nonumber\\ 
&\equiv&\frac{i}{q^2-m_{\pi^\pm}^2-i\tilde{\Sigma}_{\pi^\pm}(q_0,\vec{q})} \ .
\label{TwoPiC}
\end{eqnarray}
$m_\sigma$, $m_\pi^0$ and $m_{\pi^\pm}$ are ``bare masses'' of $\sigma$ meson, neutral pion and charged pion read off by quadratic terms in Eq.~(\ref{LMeson}). $\tilde{\Sigma}_{\sigma}(q_0,\vec{q})$, $\tilde{\Sigma}_{\pi^0}(q_0,\vec{q})$, $\tilde{\Sigma}_{\pi^+}(q_0,\vec{q})$ and $\tilde{\Sigma}_{\pi^-}(q_0,\vec{q})$ refer to the self-energies for the corresponding mesons, respectively, which are diagrammatically indicated in Fig.~\ref{fig:NNstLoops}. In these diagrams, the electric charge must be conserved at each vertex so that the intermediate particles differ among $\tilde{\Sigma}_{\pi^0}(q_0,\vec{q})$, $\tilde{\Sigma}_{\pi^+}(q_0,\vec{q})$ and $\tilde{\Sigma}_{\pi^-}(q_0,\vec{q})$. 

In the above procedure, the preservation of chiral symmetry is manifest since two-point functions for the mesons are derived from the effective action around the true ground state directly~\cite{Kapusta:2006pm}. In fact, we can confirm pion becomes massless when the explicit braking term which is proportional to $\bar{m}\epsilon$ is switched off: $\tilde{G}_{\pi^0}^{-1}(0,\vec{0}) = \tilde{G}_{\pi^+}^{-1}(0,\vec{0})= \tilde{G}_{\pi^-}^{-1}(0,\vec{0}) = 0$. 
By employing these two-point functions, the self-energy of $\bar{D}_0^*$ $(0^+)$ meson in (asymmetric) nuclear matter can be calculated as will be shown in Sec.~\ref{sec:Results}.

\section{Effective Lagrangian for $\bar{D}$ mesons}
\label{sec:HMET}
In this section, we show a derivation of a Lagrangian for ``$\bar{D}$ mesons''~\footnote{In this section, we use ``$\bar{D}$ mesons'' for referring to $\bar{D}$ ($0^-$), $\bar{D}^*$ ($1^-$), $\bar{D}_0^*$ ($0^+$) and $\bar{D}_1$ ($1^+$) mesons collectively.}. In the present study, we utilize the Heavy Quark Spin Symmetry (HQSS) and an idea of ``chiral partner structure'' for ``$\bar{D}$ mesons'' for constructing the Lagrangian. The masses of ``$\bar{D}$ mesons'' are large compared to $\Lambda_{\rm QCD}$, so that it is possible to treat ``$\bar{D}$ mesons'' which carry different spins (e.g., $\bar{D}$ ($0^-$) and $\bar{D}^*$ ($1^-$), or $\bar{D}_0^*$ ($0^+$) and $\bar{D}_1$ ($1^+$)) equivalently. In the context of chiral partner structure, a mass splitting between positive-parity meson and negative-parity meson is generated by the spontaneous breakdown of chiral symmetry. Hence, the masses of these mesons are degenerated in the chiral restoration point at tree level.

To begin with, let us introduce ``heavy-light meson fields'' $H_L$ and $H_R$. The quark contents of $H_L$ ($H_R$) is $H_L\sim c\bar{q}_L$ ($H_R \sim c\bar{q}_R$), such that $H_L$ and $H_R$ transform under the $SU(2)_L \times SU(2)_R$ chiral transformation as
\begin{eqnarray}
H_L \to H_L g_L^\dagger\ ,\ \ H_R \to H_Lg_R^\dagger\ .
\end{eqnarray}
In terms of $SU(2)_S$ heavy quark spin transformation, $H_L$ and $H_R$ transform as
\begin{eqnarray}
H_L \to S H_L\ ,\ H_R \to S H_R \ \ \ (S \in SU(2)_S)\ .
\end{eqnarray}
When we employ a picture that the motion of heavy-light meson is governed by the heavy quark, and the light quark is surrounding it like a cloud (Brown Muck picture), the kinetic terms of $H_L$ and $H_R$ can be determined by the heavy quark solely. Then, a Lagrangian for ``heavy-light meson fields'' invariant under $SU(2)_L\times SU(2)_R$ chiral transformation, $SU(2)_S$ heavy quark spin transformation and parity can be obtained as
\begin{eqnarray}
{\cal L}_{\mathrm{HMET}} &=& {\rm tr}[H_L(iv\cdot D)\bar{H}_L]+{\rm tr}[H_R(iv\cdot D)\bar{H}_R] \nonumber\\
&+& \frac{\Delta_m}{2f_{\pi}}{\rm tr}[H_LM\bar{H}_R \!+\! H_RM^{\dagger}\bar{H}_L] \nonumber\\
&+& i\frac{g_A}{2f_\pi}{\rm tr}[H_R\gamma_5\Slash{\partial} M^\dagger \bar{H}_L - H_L\gamma_5\Slash{\partial}M\bar{H}_R] \nonumber\\
&+&\frac{b_{\rho}}{2}{\rm tr}[H_L(\xi_L^\dagger v\cdot\hat{\alpha}_{\parallel}\xi_L)\bar{H}_L+H_R(\xi_R^\dagger v\cdot\hat{\alpha}_{\parallel}\xi_R)\bar{H}_R] \nonumber\\
&+& \frac{b_\omega}{2}{\rm Tr}[\hat{\alpha}_{\parallel\mu}]{\rm tr}[H_L v^\mu\bar{H}_L +H_R v^\mu\bar{H}_R]\ . \nonumber\\\label{HMETLagrangian} 
\end{eqnarray}

The kinetic term takes the form of the heavy quark effective theory, and the covariant derivatives are defined by $D_\mu \bar{H}_{L} = \partial_\mu\bar{H}_L+i\bar{H}_L{\cal L}_\mu$, $D_\mu \bar{H}_{R} = \partial_\mu\bar{H}_R+i\bar{H}_R{\cal R}_\mu$. $M$ and $\hat{\alpha}_{\parallel\mu}$ are provided by Eqs.~(\ref{MDeco}) and~(\ref{MC1Form}). $v^\mu$ is the velocity of heavy-light meson, and $\Delta_m$, $g_A$, $b_\rho$ and $b_\omega$ are real parameters. The ``tr'' in Eq.~(\ref{HMETLagrangian}) stands for the trace with respect to Dirac index. The third line is added so as to reproduce the decay of $D^* \to D\pi$. The last two lines are added to incorporate $\omega$ meson and $\rho$ meson into the ``$\bar{D}$ mesons'' system by a similar manner employed in Eq.~(\ref{NVInt}) with HLS. The effective theory defined by the Lagrangian in Eq.~(\ref{HMETLagrangian}) is referred to as the heavy meson effective theory (HMET).

$H_L$ and $H_R$ bases are convenient for constructing a Lagrangian in terms of chiral symmetry. These fields, however, are not corresponding to the physical states, i.e., the parity eigenstates. When we denote $G$ as the parity-even state while $H$ as the parity-odd state, $G$ and $H$ are related to $H_L$ and $H_R$ by the following relations:
\begin{eqnarray}
H_L &=& \frac{1}{\sqrt{2}}\left[G+iH\gamma_5\right] \ , 
\\
H_R &=& \frac{1}{\sqrt{2}}\left[G-iH\gamma_5\right] \ . 
\end{eqnarray}
As stated above, thanks to the HQSS, we can treat heavy-light mesons which carry different spins equivalently. Hence, $G$ can include $(D_0^*,D_1) =(0^+,1^+)$ while $H$ can include $(D,D^*)=(0^-,1^-)$ as 
\begin{eqnarray}
H = \frac{1+\Slash{v}}{2}[i\gamma_5D_v+\Slash{D}_v^*] \ ,
\label{PhysicalStates1} 
\\  \nonumber\\
G = \frac{1+\Slash{v}}{2}[D_{0v}^*-i\Slash{D}_{1v}\gamma_5]\ .
\label{PhysicalStates2}
\end{eqnarray}
In these equations, the subscript $v$ represents that these $D$ meson fields are defined within the HMET. In terms of $H$ and $G$, the effective Lagrangian in Eq.~(\ref{HMETLagrangian}) is rewritten into
\begin{eqnarray}
&&{\cal L}_{\rm HMET}  \nonumber\\
&=& {\rm tr}[G(iv\cdot D)\bar{G}]-{\rm tr}[H(iv\cdot D)\bar{H}] \nonumber\\
&-& i{\rm tr}[G v\cdot {\cal A} \gamma_5\bar{H}+Hv\cdot{\cal A}\gamma_5 \bar{G}] \nonumber\\
&+& \frac{\Delta_m}{4f_\pi}{\rm tr}\big[G(M+M^\dagger)\bar{G}+H(M+M^\dagger)\bar{H} \Big]\nonumber\\
&-& \frac{\Delta_m}{4f_\pi}{\rm tr}\Big[iG(M-M^\dagger)\gamma_5\bar{H}-iH(M-M^\dagger)\gamma_5\bar{G}\big] \nonumber\\ \nonumber\\
&+&i\frac{g_A}{4f_\pi}{\rm tr}\big[G\gamma_5(\Slash{\partial}M^\dagger-\Slash{\partial}M)\bar{G}-H\gamma_5(\Slash{\partial}M^\dagger-\Slash{\partial}M)\bar{H} \Big]\nonumber\\
&-&i\frac{g_A}{4f_\pi}{\rm tr}\big[iG(\Slash{\partial}M^\dagger+\Slash{\partial}M)\bar{H} +iH(\Slash{\partial}M^\dagger+\Slash{\partial}M)\bar{G}\Big] \nonumber\\
&+& \frac{b_\rho}{2} {\rm tr}[Gv\cdot\xi_\parallel^+\bar{G}-Hv\cdot\xi_\parallel^+\bar{H}] \nonumber\\
&+&i\frac{b_\rho}{2}{\rm tr}[Gv\cdot\xi_\parallel^-\gamma_5\bar{H}+Hv\cdot\xi_\parallel^-\gamma_5\bar{G}] \nonumber\\
&+& \frac{b_\omega}{2}{\rm Tr}[\hat{\alpha}_{\parallel\mu}]{\rm tr}[Gv^\mu\bar{G} - H v^\mu\bar{H}]\ ,\label{HMETLagrangianHG}
\end{eqnarray}
where we have defined $\bar{H}=\gamma^0 H^\dagger\gamma^0$, $\bar{G} = \gamma^0G^\dagger\gamma^0$, $D_\mu\bar{G}= \partial_\mu\bar{G}-i{\cal V}_\mu \bar{G}$, $D_\mu\bar{H}=\partial_\mu\bar{H}-i{\cal V}_\mu\bar{H}$ with ${\cal V}_\mu =\frac{{\cal R}_\mu+{\cal L}_\mu}{2}$, ${\cal A}_\mu=\frac{{\cal R}_\mu-{\cal L}_\mu}{2}$, and $\xi_{\parallel\mu}^{\pm}=\xi_R^\dagger\hat{\alpha}_{\parallel\mu}\xi_R\pm\xi_L^\dagger\hat{\alpha}_{\parallel\mu}\xi_L$.

By taking the charge conjugation and inserting Eqs.~(\ref{PhysicalStates1}) and~(\ref{PhysicalStates2}), we can get the effective Lagrangian for ``$\bar{D}$ mesons'' within the HMET. For later convenience, we shall derive the Lagrangian for ``$\bar{D}$ mesons'' in a relativistic form from the HMET Lagrangian~(\ref{HMETLagrangianHG}). For example, $\bar{D}$ $(0^-)$ field in the relativistic form (simply denoted by $\bar{D}$) is provided by
\begin{eqnarray}
\bar{D} = \frac{1}{\sqrt{m}}{\rm e}^{-imv\cdot x}\bar{D}_v \ ,
\end{eqnarray}
where $m$ is a parameter of dimension [mass$^1$] ($m\gg\Lambda_{\rm QCD}$). 
Therefore, the Lagrangian for ``$\bar{D}$ mesons'' based on the HQSS and the chiral partner structure in the relativistic form is of the form
\begin{widetext}
\begin{eqnarray}
{\cal L} &=& ({\cal D}^{\mu}\bar{D}_0)^{* \dagger}{\cal D}_{\mu}\bar{D}_0^*-m^2\bar{D}_0^{*\dagger}\bar{D}_0^* -m\Delta_{\bar{D}_0^*}\bar{D}_0^{*\dagger}\bar{D}_0^*-({\cal D}_{\mu}\bar{D}_{1\nu})^\dagger{\cal D}^{\mu}\bar{D}_1^{\nu}+({\cal D}_{\mu}\bar{D}_{1\nu})^\dagger {\cal D}^{\nu}\bar{D}_1^{\mu}+m^2\bar{D}_{1\mu}^\dagger\bar{D}_1^{\mu}+m\Delta_{\bar{D}_1}\bar{D}_{1\mu}^\dagger\bar{D}_1^{\mu} \nonumber\\
&+& ({\cal D}^{\mu}\bar{D})^{\dagger}{\cal D}_{\mu}\bar{D}-m^2\bar{D}^{\dagger}\bar{D}-m\Delta_{\bar{D}}\bar{D}^\dagger\bar{D}-({\cal D}_{\mu}\bar{D}_\nu^{*})^\dagger{\cal D}^{\mu}\bar{D}^{*\nu}+({\cal D}_{\mu}\bar{D}_\nu^{*})^\dagger{\cal D}^{\nu}\bar{D}^{*\mu}+m^2\bar{D}^{*\dagger}_{\mu}\bar{D}^{*\mu}+m\Delta_{\bar{D}^*}\bar{D}^{*\dagger}\bar{D}^* \nonumber\\
&-& m\frac{\Delta_m}{2f_\pi}[\bar{D}_0^{*\dagger}(M+M^{\dagger})\bar{D}_0^{*}-\bar{D}_{1\mu}^\dagger(M+M^{\dagger})\bar{D}_1^{\mu}-\bar{D}^\dagger(M+M^{\dagger})\bar{D}+\bar{D}_{\mu}^{*\dagger}(M+M^{\dagger})\bar{D}^{*\mu}] \nonumber\\
&-& m\frac{\Delta_m}{2f_\pi}[\bar{D}_0^{*\dagger}(M-M^{\dagger})\bar{D}-\bar{D}_{1\mu}^\dagger(M-M^{\dagger})\bar{D}^{*\mu}-\bar{D}^\dagger(M-M^{\dagger})\bar{D}_0^{*}+\bar{D}_{\mu}^{*\dagger }(M-M^{\dagger})\bar{D}_1^{\mu}]\nonumber\\ 
&-& m\frac{g_A}{2f_\pi}[\bar{D}_1^{\mu}(\partial_{\mu}M^{\dagger}-\partial_{\mu}M)\bar{D}_0^{*\dagger}-\bar{D}_0^{*}(\partial_{\mu}M^{\dagger}-\partial_{\mu}M)\bar{D}_1^{\dagger\mu}-\frac{1}{m}\epsilon^{\mu\nu\rho\sigma}\bar{D}_{1\mu}(\partial_{\nu}M^{\dagger}-\partial_{\nu}M)i\partial_{\sigma}\bar{D}_{1\rho}^{\dagger}] \nonumber\\
&+& m\frac{g_A}{2f_\pi}[\bar{D}^{*\mu}(\partial_{\mu}M^{\dagger}-\partial_{\mu}M)\bar{D}^{\dagger}-\bar{D}(\partial_{\mu}M^{\dagger}-\partial_{\mu}M)\bar{D}^{*\dagger\mu}-\frac{1}{m}\epsilon^{\mu\nu\rho\sigma}\bar{D}_{\mu}^*(\partial_{\nu}M^{\dagger}-\partial_{\nu}M)i\partial_{\sigma}\bar{D}_{\rho}^{*\dagger}] \nonumber\\
&+& m\frac{g_A}{2f_\pi}[\bar{D}_1^{\mu}(\partial_{\mu}M^{\dagger}+\partial_{\mu}M)\bar{D}^{\dagger}+\bar{D}(\partial_{\mu}M^{\dagger}+\partial_{\mu}M)\bar{D}_1^{\dagger\mu}]\nonumber\\
&-& m\frac{g_A}{2f_\pi}[\bar{D}_0^*(\partial_{\mu}M^{\dagger}+\partial_{\mu}M)\bar{D}^{*\dagger\mu}+\bar{D}^{*\mu}(\partial_{\mu}M^{\dagger}+\partial_{\mu}M)\bar{D}_0^{*\dagger}]\nonumber\\
&-& \frac{g_A}{2f_\pi}[\epsilon^{\mu\nu\rho\sigma}\bar{D}_{1\nu}(\partial_{\rho}M^{\dagger}+\partial_{\rho}M)i\partial_{\sigma}\bar{D}_{\mu}^{*\dagger}+\epsilon^{\mu\nu\rho\sigma}\bar{D}_{\mu}^*(\partial_{\rho}M^{\dagger}+\partial_{\rho}M)i\partial_{\sigma}\bar{D}_{1\nu}^{\dagger}] \nonumber\\
&-& b_\rho[\bar{D}_0^{*\dagger}\xi_{\parallel}^+\cdot i{\cal D}\bar{D}_0^*-\bar{D}_1^{\mu\dagger}\xi_\parallel^+\cdot i{\cal D}\bar{D}_{1\mu}-\bar{D}\xi_\parallel^+\cdot i{\cal D}\bar{D}+\bar{D}_\mu^{*\dagger}\xi_\parallel^+\cdot i{\cal D}\bar{D}^{*\mu}] \nonumber\\
&+&b_\rho[\bar{D}^\dagger \xi_\parallel^-\cdot i{\cal D}\bar{D}_0^*-\bar{D}_\mu^{*\dagger}\xi_\parallel^-\cdot i{\cal D}\bar{D}_1^\mu+\bar{D}_0^{*\dagger}\xi_\parallel^-\cdot i{\cal D}\bar{D}-\bar{D}_1^{\mu\dagger}\xi_\parallel^-\cdot i{\cal D}\bar{D}_\mu^*] \nonumber\\
&+& b_\omega{\rm Tr}[\hat{\alpha}_{\parallel\mu}](\bar{D}_0^{*\dagger}i{\cal D}^\mu\bar{D}_0^*-\bar{D}_{1\nu}^\dagger i{\cal D}^\mu\bar{D}_1^\nu+\bar{D}^\dagger i{\cal D}^\mu\bar{D}-\bar{D}^{*\dagger}_\nu i{\cal D}^\mu\bar{D}^{*\nu}) \ , \label{StartingLagrangian}
\end{eqnarray}
\end{widetext}
where we have added the small violations of HQSS for the masses of ``$\bar{D}$ mesons'' by $\Delta_{\bar{D}}$, $\Delta_{\bar{D}^*}$, $\Delta_{\bar{D}_0^*}$ and $\Delta_{\bar{D}_1}$. The covariant derivatives for ``$\bar{D}$ mesons'' are given by ${\cal D}_\mu = \partial_\mu-i{\cal V}_\mu$~\footnote{In obtaining Eq.~(\ref{StartingLagrangian}), we have assumed ${\cal V}_\mu \sim {\cal O}(\Lambda_{\rm QCD})$ and neglected higher order correspondences with Eq.~(\ref{HMETLagrangianHG}).}. By replacing as $\omega_\mu \to\bar{\omega}\delta_{\mu0}$, $\rho_\mu^a \to \bar{\rho}\delta_{\mu0}\delta^{a3}$ and ${\cal L}_\mu, {\cal R}_\mu \to (\frac{1}{3}\mu_B+\mu_I\tau^3)\delta_{0\mu}$ and redefining $\bar{\rho}$ and $\bar{\omega}$ properly as done in Sec.~\ref{sec:PDModel}, we find an effective chemical potential for ``$\bar{D}$ mesons'' of $\mu_{\bar{D}}^* = (\frac{1}{3}\mu_B-g_{\omega DD}\bar{\omega})+(\mu_I-g_{\rho DD}\bar{\rho})\tau^3$ ($g_{\omega DD} = \frac{b_\omega g_\omega-b_\rho g_\omega}{2}, g_{\rho DD}=-\frac{b_\rho g_\rho}{2}$). Note that although the effective chemical potential for ``$\bar{D}$ mesons'' $\mu_{\bar{D}}^*$ is provided, ``$\bar{D}$ mesons'' do not form Fermi seas since the masses of them are sufficiently large.

Under the spontaneous breakdown of chiral symmetry, $\sigma$ meson acquires its VEV ($\bar{\sigma}$) and the masses of ``$\bar{D}$ mesons'' are provided as
\begin{eqnarray}
M_{\bar{D}} &=& m-\frac{\Delta_m}{2f_\pi}\bar{\sigma}-\frac{\Delta_{\bar{D}}}{2}\ , \nonumber\\
M_{\bar{D}^*} &=& m-\frac{\Delta_m}{2f_\pi}\bar{\sigma}+\frac{\Delta_{\bar{D}^*}}{2}\ , \nonumber\\
M_{\bar{D}_0^*} &=& m+\frac{\Delta_m}{2f_\pi}\bar{\sigma}-\frac{\Delta_{\bar{D}_0^*}}{2}\ , \nonumber\\
M_{\bar{D}_1} &=& m+\frac{\Delta_m}{2f_\pi}\bar{\sigma}+\frac{\Delta_{\bar{D}_1}}{2}\ . \label{MassFormula}
\end{eqnarray}
The parameters $\Delta_m$, $m$, $\Delta_{\bar{D}}$, $\Delta_{\bar{D}^*}$, $\Delta_{\bar{D}_0^*}$ and $\Delta_{\bar{D}_1}$ are fixed as $m=2190$ MeV, $\Delta_m=430$ MeV, $\Delta_{\bar{D}}=202$ MeV, $\Delta_{\bar{D}^*}=80$ MeV, $\Delta_{\bar{D}_0^*}=164$ MeV and $\Delta_{\bar{D}_1}=54$ MeV so as to reproduce the observed masses of ``$\bar{D}$ mesons'': $ \accentset{\circ}{M}_{\bar{D}} = 1869$ MeV, $\accentset{\circ}{M}_{\bar{D}^*} = 2010$ MeV, $\accentset{\circ}{M}_{\bar{D}_0^*} = 2318$ MeV and $\accentset{\circ}{M}_{\bar{D}_1} = 2427$ MeV as done in Ref.~\cite{Suenaga:2017deu}. In this reference, $m$ is determined by the average value of the (spin-averaged) $G$-doublet mass and $H$-doublet mass, while $\Delta_m$ is determined by the mass difference between them. From the mass formulae in Eq.~(\ref{MassFormula}), one can easily confirm that at the chiral restoration point with the HQSS limit, all masses coincide: $M_{\bar{D}} =M_{\bar{D}^*}=M_{\bar{D}_0^*}=M_{\bar{D}_1}=m$, which shows the feature of chiral partner structure with HQSS. The parameter $g_A$ is fixed by the decay of $D^*\to D\pi$ which leads to $|g_A|=0.50$ as already mentioned.

\begin{figure}[htbp]
\centering
\includegraphics*[scale=0.4]{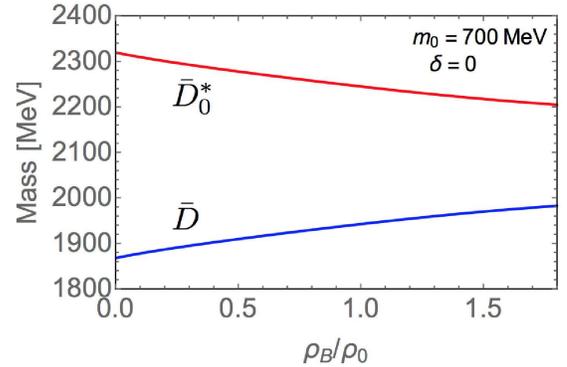}
\caption{(color online) Masses of $\bar{D}$ meson (blue curve) and $\bar{D}_0^*$ meson (red curve) at the mean field level in nuclear matter with $m_0=700$ MeV and $\delta=0$. In this level, the mass of $\bar{D}$ meson increases while that of $\bar{D}_0^*$ meson decreases which shows the characteristic feature of chiral partner structure with partial restoration of chiral symmetry in nuclear matter.}
\label{fig:DMassPlot}
\end{figure}

We plot a density dependence of masses of $\bar{D}$ meson and $\bar{D}_0^*$ meson at mean field level described by Eq.~(\ref{MassFormula}) with $m_0=700$ MeV and $\delta=0$ in Fig.~\ref{fig:DMassPlot}. The density dependence of $\bar{\sigma}$ is provided by Eq.~(\ref{StationarySigma}) (and plotted in Fig.~\ref{fig:VEV700}). The blue curve is the mass of $\bar{D}$ meson while red curve is the mass of $\bar{D}_0^*$ meson. In the mean field level, the mass of $\bar{D}$ meson increases while that of $\bar{D}_0^*$ meson decreases which shows the characteristic feature of chiral partner structure with partial restoration of chiral symmetry in nuclear matter. Even when we include asymmetry by taking $\delta\neq0$, the result does not change significantly since the value of $\bar{\sigma}$ shows only a small change.

\section{Calculations and results}
\label{sec:Results}
\begin{figure*}[htbp]
\centering
\includegraphics*[scale=0.4]{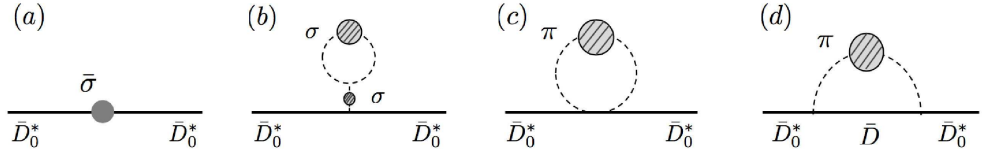}
\caption{Relevant self-energies for $\bar{D}_0^*$ ($0^+$) meson in nuclear matter. The blobs in the two-point functions for $\sigma$ and pion indicate infinite sums of self-energies in Fig.~\ref{fig:NNstLoops} since the two-point functions are derived by Eqs.~(\ref{TwoSigma})-~(\ref{TwoPiC}). The diagram $(a)$ is the mean field contribution, and $(b)$ and $(c)$ are corresponding to the Hartree-type one-loop modifications. These three diagrams are not non-local and not momentum dependent. The diagram $(d)$ is the dominant Fock-type one-loop contribution to the self energy such that this diagram generates a momentum dependence and an imaginary part which plays a significant role in plotting a spectral function for $\bar{D}_0^*$ meson.}
\label{fig:DSelfEnergy}
\end{figure*}

In this section, we calculate a self energy for $\bar{D}_0^*$ ($0^+$) meson in symmetric and asymmetric nuclear matter and show resultant spectral functions for $\bar{D}_0^*$ meson. In terms of the self energy, we evaluate Hartree-type and Fock-type one-loop diagrams in addition to the mean field modification as depicted in Fig.~\ref{fig:DSelfEnergy}. 
In Sec.~\ref{sec:SymmetricMatter}, we show results in symmetric nuclear matter at several densities with $m_0=500$ MeV and $m_0=700$ MeV. In Sec.~\ref{sec:AsymmetricMatter}, we show results in asymmetric nuclear matter with $m_0=500$ MeV and $m_0=700$ MeV. The detailed calculation of the self energy is given in Appendix.~\ref{sec:OneLoopS} and Appendix.~\ref{sec:OneLoopAS}.

\subsection{Spectral function for $\bar{D}_0^*$ meson in symmetric nuclear matter}
\label{sec:SymmetricMatter}
In this subsection, we compute the self energy and spectral function for $\bar{D}_0^*$ $(0^+)$ meson in symmetric nuclear matter. The relevant diagrams are depicted in Fig.~\ref{fig:DSelfEnergy}. In these diagrams, the blobs in the two-point functions for $\sigma$ and pion indicate infinite sums of self-energies in Fig.~\ref{fig:NNstLoops} since the two-point functions are derived by Eqs.~(\ref{TwoSigma})-~(\ref{TwoPiC}). The diagram $(a)$ is the mean field contribution, and $(b)$ and $(c)$ are corresponding to the Hartree-type one-loop modifications. These three diagrams are not non-local and not momentum dependent. The diagram $(d)$ is the Fock-type one-loop contribution to the self energy which generates a momentum dependence to the self energy. Beside, this diagram provides an imaginary part to the self energy and plays a significant role in plotting the spectral function for $\bar{D}_0^*$ meson. Note that although the Lagrangian~(\ref{StartingLagrangian}) yields the other Fock-type one-loop diagrams, the diagram in Fig.~\ref{fig:DSelfEnergy} ($d$) governs the results. Then, we compute only Fig.~\ref{fig:DSelfEnergy} ($d$) as a good approximation.

The mean field contribution in Fig.~\ref{fig:DSelfEnergy} ($a$) is easily calculated by the mass formula in Eq.~(\ref{MassFormula}) with $\bar{\sigma}$ being the solution of the gap equation~(\ref{StationarySigma}). In other words, the mean field effect is automatically included when we use Eq.~(\ref{MassFormula}) as the $\bar{D}_0^*$ meson ``bare mass''.

The retarded self-energy generated by the Hartree-type contributions in Fig.~\ref{fig:DSelfEnergy} ($b$) and ($c$) are calculated as (For the detail, see Appendix.~\ref{sec:OneLoopS})
\begin{eqnarray}
&&\tilde{\Sigma}_{\bar{D}_0^*(H)}^R \nonumber\\
&=&-\frac{3m\Delta_m}{4\pi^3f_\pi \tilde{m}_\sigma^2}\left(\lambda\bar{\sigma}-\frac{10}{3}\lambda_6\bar{\sigma}^3\right) \int_0^\infty dk_0\int_0^\infty d|\vec{k}|\, |\vec{k}|^2\nonumber\\ \nonumber\\
&&\times\left(F(|\vec{k}|,\Lambda)\right)^2\left\{\rho_\sigma(k_0,\vec{k})-\accentset{\circ}{\rho}_\sigma(k_0,\vec{k})\right\} \nonumber\\
&-&3\frac{m\Delta_m}{8\pi^3f_\pi\bar{\sigma}}\int_0^\infty dk_0\int_0^\infty d|\vec{k}|\, |\vec{k}|^2\nonumber\\
&&\times\left(F(|\vec{k}|,\Lambda)\right)^2\left\{\rho_{\pi}(k_0,\vec{k})-\accentset{\circ}{\rho}_{\pi}(k_0,\vec{k})\right\}\ , \label{Hartree}
\end{eqnarray}
where we have denoted $\pi$ for $\pi^0$, $\pi^+$ and $\pi^-$ collectively since there is no differences among them due to the isospin symmetry. $\tilde{m}_\sigma^2$ is the (squared) mass of $\sigma$ meson in nuclear matter defined by a pole of the two-point function with vanishing three-momentum in Eq.~(\ref{TwoSigma}). $F(|\vec{k}|,\Lambda)$ is a form factor inserted in order to take into account the finite size of hadrons which takes the form of
\begin{eqnarray}
F(|\vec{k}|,\Lambda) = \frac{\Lambda^2}{|\vec{k}|^2+\Lambda^2}\ .
\end{eqnarray}
The value of cutoff $\Lambda$ is taken to be $\Lambda=400$ MeV which is slightly higher than the Fermi momentum in the present study. $\rho_\sigma(k_0,\vec{k})$ ($\rho_{\pi}(k_0,\vec{k})$) and $\accentset{\circ}{\rho}_\sigma(k_0,\vec{k})$ ($\accentset{\circ}{\rho}_{\pi}(k_0,\vec{k})$) are the spectral functions for $\sigma$ meson (pion) in nuclear matter and in the vacuum, respectively, given by
\begin{eqnarray}
\rho_{\sigma(\pi)}(k_0,\vec{k}) &=& -2{\rm Im}\left[\frac{1}{k^2-m^2_{\sigma(\pi)}-\tilde{\Sigma}^R_{\sigma(\pi)}(k_0,\vec{k})}\right] \nonumber\\ \nonumber\\
\accentset{\circ}{\rho}_{\sigma(\pi)}(k_0,\vec{k}) &=&2\pi\epsilon(k_0)\delta(k^2-m^2_{\sigma(\pi)})\ .
\end{eqnarray}
In this equations, $\epsilon(k_0)$ is the sign-function defined by $\epsilon(k_0) = +1 (-1)$ for $k_0>0$ ($k_0<0$), and $\tilde{\Sigma}^R_{\sigma(\pi)}(k_0,\vec{k})$ is the retarded self-energy related to the self-energy $\tilde{\Sigma}_{\sigma(\pi)}(k_0,\vec{k})$ defined in Eqs.~(\ref{TwoSigma}) -~(\ref{TwoPiC}) by the following relations:
\begin{eqnarray}
{\rm Re}\tilde{\Sigma}^R_{\sigma(\pi)}(k_0,\vec{k}) &=& {\rm Re}\left(i\tilde{\Sigma}_{\sigma(\pi)}(k_0,\vec{k})\right) \nonumber\\
{\rm Im}\tilde{\Sigma}^R_{\sigma(\pi)}(k_0,\vec{k}) &=& \epsilon(k_0){\rm Im}\left(i\tilde{\Sigma}_{\sigma(\pi)}(k_0,\vec{k})\right)\ .
\end{eqnarray}
We should note that we have subtracted the spectral function in the vacuum in Eq.~(\ref{Hartree}) so as to renormalize the one-loop correction as $\tilde{\Sigma}_{\bar{D}_0^*(H)}=0$ in the vacuum.

Next, we show a calculation of the Fock-type corrections to the self-energy for $\bar{D}_0^*$ meson in Fig.~\ref{fig:DSelfEnergy} ($d$). We employ so-called ``spectral representation method~\cite{TFT}''  for the calculation. The detail and a concrete demonstration of this method is provided in Sec. IV B in Ref.~\cite{Suenaga:2017deu}. According to this method, the imaginary part of the retarded self-energy in Fig.~\ref{fig:DSelfEnergy} ($d$) is obtained as
\begin{widetext}
\begin{eqnarray}
{\rm Im}\tilde{\Sigma}^R_{\bar{D}_0^*(F)}(q_0,\vec{q}) &=& -\frac{3}{2}\left(\frac{m\Delta_m}{f_\pi}\right)^2\int\frac{d^3k}{(2\pi)^3}\left( F(|\vec{k}|,\Lambda)\right)^2 \nonumber\\
&& \times \frac{1}{2E_k^{\bar{D}}} \Bigg\{\Big(\theta(q_0-E_k^{\bar{D}})\rho_{\pi}(q_0-E_k^{\bar{D}},\vec{q}-\vec{k}) +\theta(-q_0-E_k^{\bar{D}})\rho_{\pi}(q_0+E_k^{\bar{D}},\vec{q}-\vec{k})\Bigg\}\ ,  \label{ImD0st}
\end{eqnarray}
\end{widetext}
where we have defined $E_k^{\bar{D}} = \sqrt{|\vec{k}|^2+M_{\bar{D}}^2}$. $M_{\bar{D}}$ is the ``bare mass'' of $\bar{D}$ meson defined in Eq.~(\ref{MassFormula}) since the perturbation series are defined around the true ground state determined by Eqs.~(\ref{StationarySigma}) -~(\ref{StationaryRho}). The real part of the retarded self-energy is calculated via the following subtracted dispersion relation 
\begin{eqnarray}
&&{\rm Re}\tilde{\Sigma}_{\bar{D}_0^*(F)}^R(q_0,\vec{q})\nonumber\\
&=&\frac{q_0-\accentset{\circ}{E}_q^{\bar{D}_0^*}}{\pi}  {\rm P}\int_{-\infty}^\infty dz\frac{{\rm Im}\tilde{\Sigma}_{\bar{D}_0^*(F)}^R(z,\vec{q})}{(z-q_0)(z-\accentset{\circ}{E}_q^{\bar{D}_0^*})} \ , \nonumber\\ \label{ReD0st}
\end{eqnarray}
with $\accentset{\circ}{E}_q^{\bar{D}_0^*}=\sqrt{|\vec{q}|^2+\accentset{\circ}{M}_{\bar{D}_0^*}^2}$ ($\accentset{\circ}{M}_{\bar{D}_0^*}$ is the $\bar{D}_0^*$ meson mass in the vacuum). As utilized in the calculation of Hartree-type one-loop diagrams in Eq.~(\ref{Hartree}), we have renormalized the Fock-type one-loop diagram so as to read ${\rm Re}\tilde{\Sigma}_{\bar{D}_0^*(F)}^R(\accentset{\circ}{E}_{\bar{D}_0^*},\vec{q})=0$ in the vacuum.
\begin{figure*}[htbp]
\centering
\includegraphics*[scale=0.48]{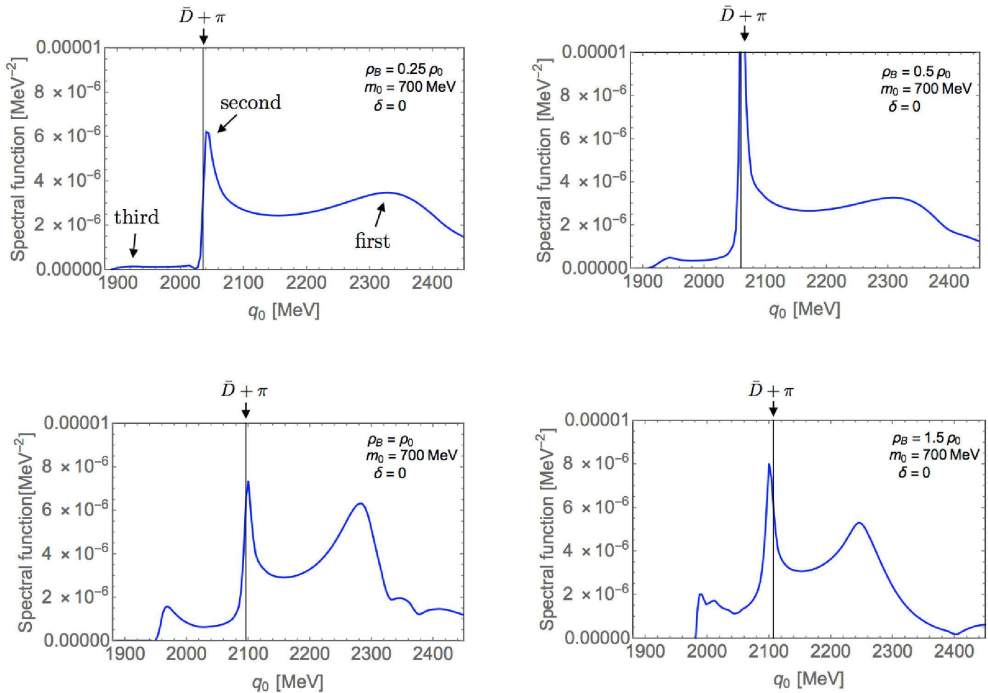}
\caption{(color online) Spectral functions for $\bar{D}_0^*$ meson at rest $\vec{q}=\vec{0}$ at several densities with $m_0=700$ MeV and $\delta=0$. The vertical black line represents the threshold of $\bar{D} + \pi$. The detail is given in the text.}
\label{fig:Delta0M0700}
\end{figure*}

From Eqs.~(\ref{ImD0st}) and~(\ref{ReD0st}), a spectral function for $\bar{D}_0^*$ meson in symmetric nuclear matter
\begin{eqnarray}
\rho_{\bar{D}_0^*}^*(q_0,\vec{q}) =-2{\rm Im}\left[ \frac{1}{q^2-M_{\bar{D}_0^*}^2-\tilde{\Sigma}^R_{\bar{D}_0^*}(q_0,\vec{q})}\right] \nonumber\\
\end{eqnarray}
with
\begin{eqnarray}
&&\tilde{\Sigma}^R_{\bar{D}_0^*}(q_0,\vec{q}) \nonumber\\
&=&\tilde{\Sigma}^R_{\bar{D}_0^*(H)}(q_0,\vec{q})+{\rm Re}\tilde{\Sigma}^R_{\bar{D}_0^*(F)}(q_0,\vec{q})+i{\rm Im}\tilde{\Sigma}^R_{\bar{D}_0^*(F)}(q_0,\vec{q}) \nonumber\\
\end{eqnarray}
is found. The resultant spectral function at rest $\vec{q}=\vec{0}$ at several densities with $m_0=700$ MeV is plotted in Fig.~\ref{fig:Delta0M0700}. As we can see, three types of bumps are found.

The first bump from right corresponds to the $\bar{D}_0^*$ meson resonance. This bump gets suppressed as the density increases from $\rho_B=0.25\rho_0 $ to $\rho_B=0.5\rho_0$ due to a collisional broadening caused by collisions with nucleons surrounding $\bar{D}_0^*$ meson. In contrast, an enhancement of this bump is observed at $\rho_B=\rho_0$ or $\rho_B=1.5\rho_0$ in comparison to the results at lower density. This revival is understood as follows: As the density increases, the chiral symmetry tends to be restored and the mass difference between $\bar{D}_0^*$ meson and $\bar{D}$ meson gets small at mean field level since we have introduced these mesons as chiral partners as indicated in Fig.~\ref{fig:DMassPlot}.  As a result, the phase space of $\bar{D}_0^* \to \bar{D}\pi$ gets narrowed even though the collisional broadening still operates, which leads to the enhancement of the $\bar{D}_0^*$ resonance. We should emphasize that this behavior is provided by the partial restoration of chiral symmetry and chiral partner structure.
\begin{figure*}[htbp]
\centering
\includegraphics*[scale=0.5]{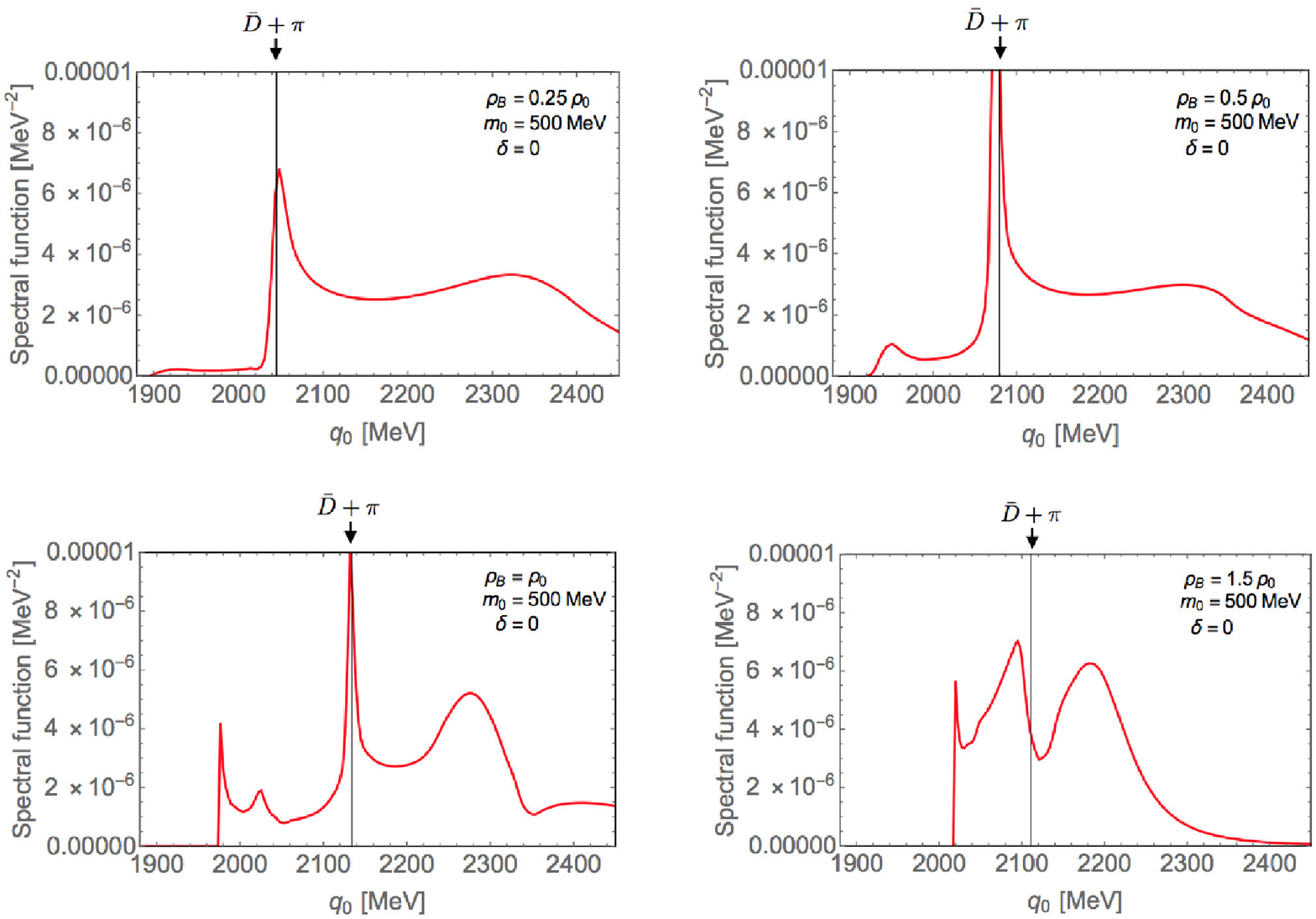}
\caption{(color online) Spectral functions for $\bar{D}_0^*$ meson at rest $\vec{q}=\vec{0}$ at several densities with $m_0=500$ MeV and $\delta=0$.}
\label{fig:Delta0M0500}
\end{figure*}

The second bump from right corresponds to a threshold enhancement. This peak stands at the threshold of $\bar{D} + \pi$ and remarkably enhanced at density. At the normal nuclear matter density, its height competes with $\bar{D}_0^*$ resonance, however, this peak is sharp. Besides, the peak position represents the mass of $\bar{D}$ meson directly at mean field level since pion mass is not changed at density compared to that of $\bar{D}$ meson. Therefore, this peak is a proper probe to explore the magnitude of partial restoration of chiral symmetry~\cite{Suenaga:2017deu}. The origin of this enhancement may be understood by a virtual state or a bound state of $\bar{D}$ meson and $\pi$, which are defined by a pole in the second Riemann sheet or the physical sheet~\cite{Hidaka:2002xv}.

The third bump from right is mainly caused by the Landau damping. This effect is corresponding to a scattering process between $\bar{D}_0^*$ meson and nucleon in medium mediated by pion ($\bar{D}_0^* +N \to \bar{D}+N$). This bump gradually grows as we access to the higher density.

In Fig.~\ref{fig:Delta0M0500}, we plot the spectral functions for $\bar{D}_0^*$ meson at rest $\vec{q}=\vec{0}$ at several densities with $m_0=500$ MeV. The qualitative tendencies are not changed by the results obtained with $m_0=700$ MeV. The partial restoration of chiral symmetry is strengthened compared to $m_0=700$ MeV, so that the density dependences of three types of bumps are more prominent.

\subsection{Spectral function for $\bar{D}_0^*$ meson in neutron-rich asymmetric nuclear matter}
\label{sec:AsymmetricMatter}
Here, we show results of spectral functions for $\bar{D}_0^*$ mesons at rest $\vec{q}=\vec{0}$ in neutron-rich asymmetric nuclear matter with $(m_0,\delta)=(700\, {\rm MeV}, -0.5)$, $(700\, {\rm MeV},-1.0)$, $(500\, {\rm MeV},-0.5)$, $(500\, {\rm MeV},-1.0)$. The baryon number density is chosen to be the normal nuclear matter density: $\rho_B=\rho_0$. We should note that neutral $\bar{D}_0^*$ meson ($\sim \bar{c}u$) is denoted by $\bar{D}_{0,u}^*$ while negatively-charged $\bar{D}_0^*$ meson ($\sim \bar{c}d$) is $\bar{D}_{0,d}^*$ in this paper.

 In asymmetric nuclear matter, it is worth evaluating the masses of $\pi^0$, $\pi^+$ and $\pi^-$ since these masses differ generally because of the violation of isospin symmetry. 
\begin{table*}[htb]
  \begin{tabular}{cc||ccc} \hline\hline 
$m_0$ (MeV) & $\delta$ & $\tilde{m}_{\pi^0}$ (MeV) & $\tilde{m}_{\pi^+}$ (MeV) & $\tilde{m}_{\pi^-}$ (MeV) \\\hline
$500$ &$0$ & $160$ & $160$ & $160$ \\
$700$ & $0$ & $154$ &$154$ & $154$  \\
$500$ & -0.5 & $160$ & $155$ & $166$ \\
$700$ & -0.5 & $154$ & $148$ & $161$  \\
$500$ & -1 & $160$ & $155$ & $169$\\
$700$ & -1 & $154$ & $143$ & $167$ 
\\ \hline \hline
  \end{tabular}
\caption{Masses of $\pi^0$, $\pi^+$ and $\pi^-$ with several choices of $m_0$ and $\delta$. Here, the masses are defined by the poles of two-point functions defined by Eqs.~(\ref{TwoPi0}) and~(\ref{TwoPiC}) with vanishing three-momentum.}
  \label{tab:PiMass}
\end{table*}
The results are listed in Table.~\ref{tab:PiMass}. Note that the masses of $\pi^0$, $\pi^+$ and $\pi^-$ in asymmetric nuclear matter must be defined by the poles of two-point functions defined by Eqs.~(\ref{TwoPi0}) and~(\ref{TwoPiC}) with vanishing three-momentum. Despite the masses are identical thanks to the isospin symmetry for $\delta=0$, they show a mass hierarchy of $\tilde{m}_{\pi^+} < \tilde{m}_{\pi^0}<\tilde{m}_{\pi^-}$ when the asymmetry is present. $\pi^-$ consists of a $d$ quark and an anti-$u$ quark such that this meson feels a repulsive force by the neutron-rich matter due to the Pauli blocking. As a results, the mass of $\pi^-$ is greater than those of $\pi^+$ and $\pi^0$. In contrast, $\pi^+$ feels an attractive force by the matter and its mass is smaller that other two pions. This mass hierarchy provides a difference between spectral functions for $\bar{D}_{0,u}^*$ and $\bar{D}_{0,d}^*$ mesons in asymmetric nuclear matter since the spectra are dominated by the one pion decay as already stated.

\begin{figure*}[htbp]
\centering
\includegraphics*[scale=0.55]{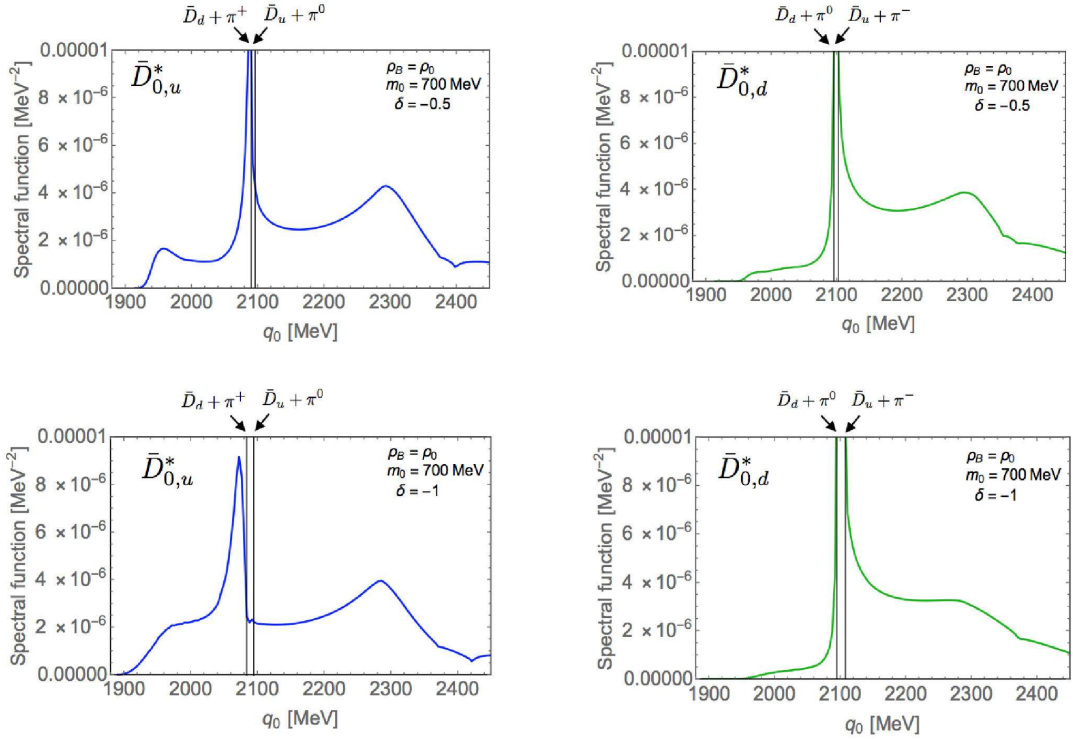}
\caption{(color online) Spectral functions for $\bar{D}_{0,u}^*$ and $\bar{D}_{0,d}^*$ mesons in asymmetric nuclear matter at rest $\vec{q}=\vec{0}$ with $m_0=700$ MeV.}
\label{fig:AsymmetricM0700}
\end{figure*}
\begin{figure*}[htbp]
\centering
\includegraphics*[scale=0.55]{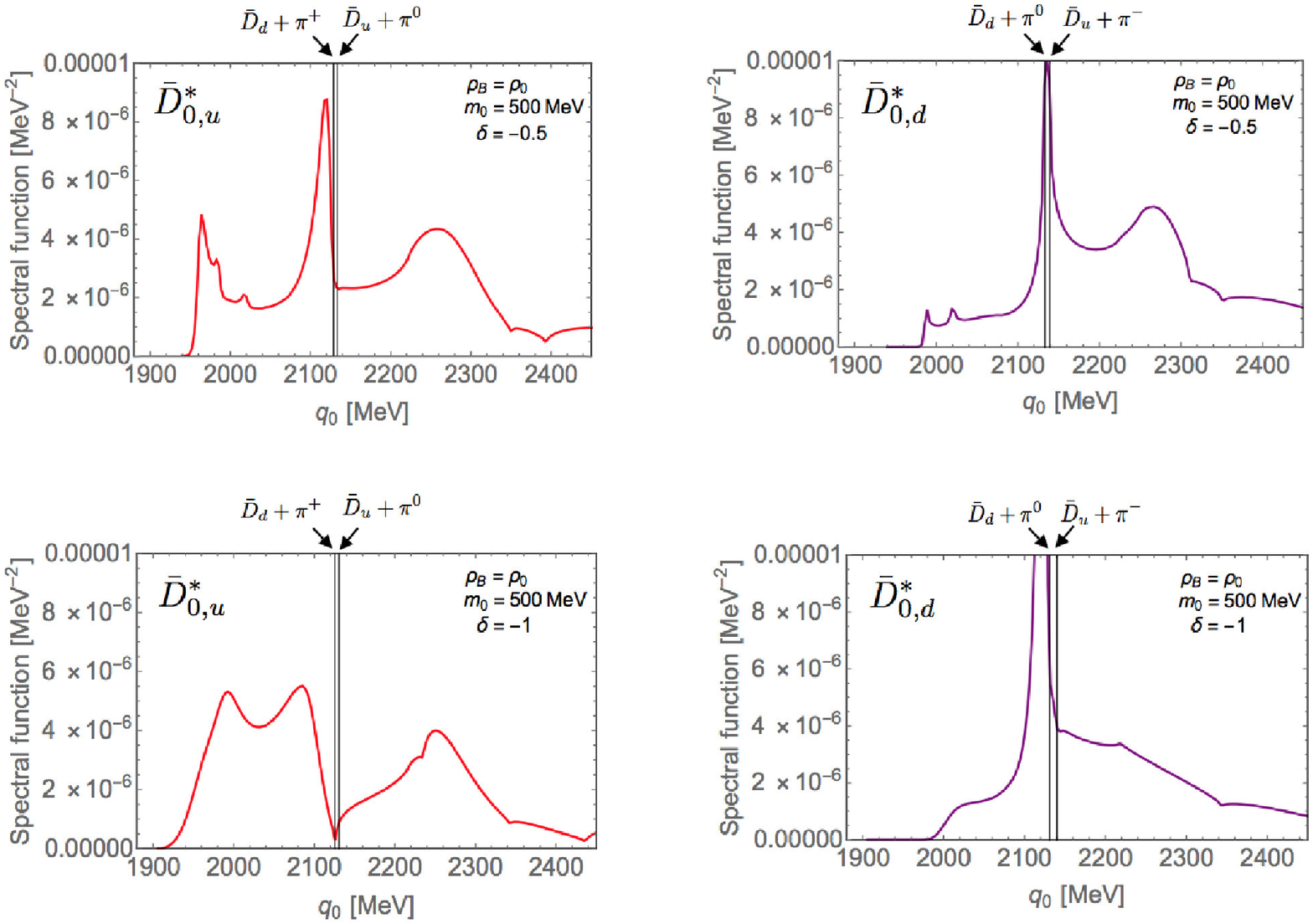}
\caption{(color online) Spectral functions for $\bar{D}_{0,u}^*$ and $\bar{D}_{0,d}^*$ mesons in asymmetric nuclear matter at rest $\vec{q}=\vec{0}$ with $m_0=500$ MeV.}
\label{fig:AsymmetricM0500}
\end{figure*}

The resultant spectral functions for $\bar{D}_{0,u}^*$ and $\bar{D}_{0,d}^*$ mesons at rest $\vec{q}=\vec{0}$ are plotted in Fig.~\ref{fig:AsymmetricM0700} and Fig.~\ref{fig:AsymmetricM0500}. In these figures, the left panels show the results for $\bar{D}_{0,u}^*$ meson and the right panels show the results for $\bar{D}_{0,d}^*$ meson. The vertical lines denote the threshold of each channel. At first sight, we find complicated structures due to the violation of isospin symmetry. However, the qualitative classification of the bumps are not changed from the results in symmetric nuclear matter, i.e., the bumps are corresponding to the resonance of $\bar{D}_{0,u}^*$ ($\bar{D}_{0,d}^*$) meson, threshold enhancement and (mainly) Landau damping.

From these figures, we can see that the threshold enhancement is remarkably enhanced as the results in symmetric nuclear matter in Sec~\ref{sec:SymmetricMatter}. In addition, these figures show this enhancement is more prominent for $\bar{D}_{0,d}^*$ meson channel. This difference is essentially caused by the difference of position of the threshold: For $\bar{D}_{0,d}^*$ meson channel, the main decay modes are $\bar{D}_{0,d}^* \to \bar{D}_d + \pi^0$ and $\bar{D}_{0,d}^*\to \bar{D}_u+\pi^-$. Since mass of $\pi^-$ is greater than those of $\pi^0$ and $\pi^+$ as indicated in Table.~\ref{tab:PiMass}, the peak position of the threshold enhancement stands at higher energy compared to $\bar{D}_{0,u}^*$ meson channel. As a result, the threshold enhancement in $\bar{D}_{0,d}^*$ meson channel gets more prominent. These results suggest it is more appropriate to focus on $\bar{D}_{0,d}^*$ meson to investigate partial restoration of chiral symmetry in (neutron-rich) asymmetric nuclear matter.

Also, We find that the resonance of $\bar{D}_{0,d}^*$ meson is slightly suppressed compared to $\bar{D}_{0,u}^*$ meson. For $\bar{D}_{0,d}^*$ meson, the Pauli blocking effect is more prominent since the matter is neutron-rich so that this suppression may be caused. We also find that the bump below the threshold (third-type bump) is also suppressed in $\bar{D}_{0,d}^*$ meson channel.

\section{Conclusion}
\label{sec:Conclusion}
In this paper, we study spectral function for $\bar{D}_0^*$ $(0^+)$ meson in isospin symmetric and neutron-rich asymmetric nuclear matter in terms of partial restoration of chiral symmetry. 

In Sec.~\ref{sec:PDModel}, we construct (asymmetric) nuclear matter by a parity doublet model in which a Fermion one-loop and mean field approximation for $\sigma$, $\rho$ and $\omega$ mesons are employed. In constructing (asymmetric) nuclear matter, we take properties at normal nuclear matter density as well as vacuum properties as inputs (input parameters are summarized in Table.~\ref{tab:Parameters}), and only $m_0$ is remained as a free parameter in the present analysis. This $m_0$ is so-called a chiral invariant mass defined by the nucleon mass at chiral restoration point. Density dependences of the mean fields of $\sigma$, $\rho$ and $\omega$ mesons are determined by stationary conditions of the thermodynamic potential. Therefore, we can access to the normal nuclear matter density self consistently. Within this model, the partial restoration of chiral symmetry in nuclear matter is found as shown in Fig.~\ref{fig:VEV700}.

In Sec.~\ref{sec:HMET}, we introduce ``$\bar{D}$ mesons'' by a chiral partner structure and heavy quark spin symmetry. In the context of chiral partner structure, a mass difference between $\bar{D}$ ($0^-$) and $\bar{D}_0^*$ ($0^+$) mesons is generated by the spontaneous breakdown of chiral symmetry. Then, it is expected that the mass difference gets small as we increase the density due to the partial restoration of chiral symmetry. This tendency is found in Fig.~\ref{fig:DMassPlot} at the mean field level. In particular, we observe the increase in mass of $\bar{D}$ meson and the decrease in $\bar{D}_0^*$ meson at finite baryon density.

In Sec.~\ref{sec:Results}, we calculate self-energy for $\bar{D}_0^*$ meson and show results of spectral functions for $\bar{D}_0^*$ meson in symmetric and neutron-rich asymmetric nuclear matter. In our calculation, we utilize a naive picture of which interactions between ``$\bar{D}$ mesons'' and (asymmetric) nuclear matter is triggered by one pion and $\sigma$ meson exchanges. The two-point functions of mediating pion and $\sigma$ meson must be derived by the effective action describing (asymmetric) nuclear matter directly in order to respect the original chiral symmetry. The self energy is evaluated by calculating a mean field correction, Hartree-type one-loop corrections, and a Fock-type one-loop correction depicted in Fig.~\ref{fig:DSelfEnergy}. The decay mode of $\bar{D}_0^*$ meson is governed by $\bar{D}_0^*\to \bar{D}\pi$ so that we include only $\bar{D}\pi$ loop into the Fock-type diagram as a proper approximation.

The resulting spectral functions in symmetric nuclear matter at rest $\vec{q}=\vec{0}$ with $m_0=700$ MeV and $m_0=500$ MeV are shown in Fig.~\ref{fig:Delta0M0700} and Fig.~\ref{fig:Delta0M0500}, respectively. We find three types of bumps. The first bump corresponds to a $\bar{D}_0^*$ resonance. This bump is suppressed around $\rho_B=0.25 \rho_0$ - $0.5 \rho_0$ due to a collisional broadening. In contrast, this bump is enhanced around $\rho_B=1.0\rho_0$ - $1.5 \rho_0$, i.e., a revival of $\bar{D}_0^*$ meson resonance is found. This is caused by a narrowing of the phase space of $\bar{D}_0^* \to \bar{D}\pi$ decay because of the partial restoration of chiral symmetry. The second bump corresponds to a threshold enhancement which stands at the threshold of $\bar{D}+\pi$ originated from a virtual or bound state of $\bar{D}$ meson and pion. This bump is remarkably enhanced at density and its peak position reflects the mass of $\bar{D}$ meson directly. Therefore, it is expected that this peak is an appropriate probe to investigate the magnitude of partial restoration of chiral symmetry in nuclear matter as well as the value of chiral invariant mass $m_0$. The third bump is mainly caused by a Landau damping.

Spectral functions for $\bar{D}_{0}^*$ meson in neutron-rich asymmetric nuclear matter at rest $\vec{q}=\vec{0}$ are shown in Fig.~\ref{fig:AsymmetricM0700} and Fig.~\ref{fig:AsymmetricM0500} ($\bar{D}_{0,u}^*$ refers to the neutral $\bar{D}_0^*$ meson while $\bar{D}_{0,d}^*$ refers to the negatively-charged $\bar{D}_0^*$ meson).  We observe that the peak position of the threshold enhancement in $\bar{D}_{0,d}^\ast$ meson channel is located at higher energy than that in $\bar{D}_{0,u}^\ast$ meson channel since we have a mass hierarchy in pion masses of $\tilde{m}_{\pi^+}<\tilde{m}_{\pi^0}<\tilde{m}_{\pi^-}$. Accordingly, the enhancement of this peak in $\bar{D}_{0,d}^\ast$ channel is more prominent compared to $\bar{D}_{0,u}^*$ channel. This fact claims that it is appropriate to focus on the spectral function for $\bar{D}_{0,d}^*$ meson when we study the partial restoration of chiral symmetry by means of anti-charmed mesons in neutron-rich asymmetric matter. We also find that the resonance of $\bar{D}_{0,d}^*$ meson and the bump below the threshold (the third bump) is suppressed compared to those of the $\bar{D}_{0,u}^*$ meson.

These results provide useful information of partial restoration of chiral symmetry for the future experiment such as the PANDA experiment at FAIR. Moreover, it is expected that the results are necessary to understand the collective behavior in low energy heavy ion collisions planed in the CBM experiment at FAIR, or the J-PARC-HI program at J-PARC.

In the following, we give discussions. In the present analysis, we take the cutoff parameter to be $\Lambda=400$ MeV which is slightly higher than the Fermi momentum in showing our results. When we change the value, the hight of the bumps may change slightly. However, the qualitative tendencies of the results are not lost, i.e., three-types of bumps are found and the threshold enhancement is remarkably enhanced. In particular, the position of $\bar{D}+\pi$ threshold does not vary even if we take another value of $\Lambda$, and our conclusion is not changed. Besides, we do not include any other Fock-type diagrams but Fig.~\ref{fig:DSelfEnergy} $(d)$ in calculating the self-energy of $\bar{D}_0^*$ meson. We confirm this approximation works well numerically.

In the present study, we include the nucleon ($N(939)$) and $N^*(1535)$ loops in the calculation of two-point functions of pion and $\sigma$ meson. It is expected, however, that contributions from $\Delta$ should also be included (e.g., $N$-$\Delta$ loops). These effects may contribute to broaden the $\bar{D}_0^*$ resonance and Landau damping. In spite of these modifications, the peak position of the threshold enhancement may show a small change and our main conclusion is stable since the position is mainly determined by the mass of $\bar{D}$ meson.

In order to confirm our results in a real experiment, it is necessary to compute observables such as a double differential cross section~ \cite{Yamagata-Sekihara:2015ebw,Shyam:2016bzq}. It is interesting to compute it with our spectral function. We leave this subject in the future work.

\acknowledgments

This work is supported partly by the Grant-in-Aid for JSPS Research Fellow No. 17J05638 (D.S.).

\appendix
\section{Calculations of one-loop diagrams in Fig.~\ref{fig:DSelfEnergy} in symmetric nuclear matter}
\label{sec:OneLoopS}
Here, we shall show a explicit calculation of the one-loop diagrams in Fig.~\ref{fig:DSelfEnergy}. First, we calculate the Hartree-type one-loop diagrams in Fig.~\ref{fig:DSelfEnergy} $(b)$ and $(c)$. By making use of the ordinary Feynman rule, the diagram in Fig.~\ref{fig:DSelfEnergy} $(b)$ yields
\begin{widetext}
\begin{eqnarray}
\tilde{\Sigma}_{{\rm Fig}. 4\, (b)} &=& -3\frac{m\Delta_m}{f_\pi \tilde{m}_\sigma^2}\left(\lambda\bar{\sigma}-\frac{10}{3}\lambda_6\bar{\sigma}^3\right)\int\frac{d^4k}{(2\pi)^4}\left(F(|\vec{k}|,\Lambda)\right)^2\left\{ \tilde{G}_\sigma(k_0,\vec{k})-\accentset{\circ}{\tilde{G}}_\sigma(k_0,\vec{k})\right\} \nonumber\\
&=& 3\frac{m\Delta_m}{f_\pi \tilde{m}_\sigma^2}\left(\lambda\bar{\sigma}-\frac{10}{3}\lambda_6\bar{\sigma}^3\right)\int\frac{d^4k}{(2\pi)^4}\left(F(|\vec{k}|,\Lambda)\right)^2{\rm Im}\left[\frac{1}{k^2-m_\sigma^2-i\tilde{\Sigma}_\sigma(k_0,\vec{k})}-\frac{1}{k^2-\accentset{\circ}{m}_\sigma^2+i\epsilon}\right] \nonumber\\
&=& -3\frac{m\Delta_m}{2f_\pi \tilde{m}_\sigma^2}\left(\lambda\bar{\sigma}-\frac{10}{3}\lambda_6\bar{\sigma}^3\right)\int\frac{d^4k}{(2\pi)^4}\left(F(|\vec{k}|,\Lambda)\right)^2\left\{\epsilon(k_0)\rho_{\sigma}(k_0,\vec{k})-\epsilon(k_0)\accentset{\circ}{\rho}_{\sigma}(k_0,\vec{k})\right\} \nonumber\\
&=&  -3\frac{m\Delta_m}{4\pi^3f_\pi \tilde{m}_\sigma^2}\left(\lambda\bar{\sigma}-\frac{10}{3}\lambda_6\bar{\sigma}^3\right)\int_0^\infty dk_0\int_0^\infty d|\vec{k}||\vec{k}|^2\left(F(|\vec{k}|,\Lambda)\right)^2\left\{\rho_{\sigma}(k_0,\vec{k})-\accentset{\circ}{\rho}_{\sigma}(k_0,\vec{k})\right\}\ . \label{Tadpole1}
\end{eqnarray}
In a similar way, the diagram in Fig.~\ref{fig:DSelfEnergy} $(c)$ is calculated:
\begin{eqnarray}
\tilde{\Sigma}_{{\rm Fig}. 4\, (c)}  &=& - 3\frac{m\Delta_m}{2f_\pi\bar{\sigma}}\int\frac{d^4k}{(2\pi)^4}\left(F(|\vec{k}|,\Lambda)\right)^2\left\{ \tilde{G}_\pi(k_0,\vec{k})-\accentset{\circ}{\tilde{G}}_\pi(k_0,\vec{k})\right\}  \nonumber\\
&=& 3\frac{m\Delta_m}{2f_\pi\bar{\sigma}}\int\frac{d^4k}{(2\pi)^4}\left(F(|\vec{k}|,\Lambda)\right)^2{\rm Im}\left[\frac{1}{k^2-m_\pi^2-i\tilde{\Sigma}_\pi(k_0,\vec{k})}-\frac{1}{k^2-\accentset{\circ}{m}_\pi^2+i\epsilon}\right] \nonumber\\
&=& -3\frac{m\Delta_m}{4f_\pi\bar{\sigma}}\int\frac{d^4k}{(2\pi)^4}\left(F(|\vec{k}|,\Lambda)\right)^2\left\{\epsilon(k_0)\rho_{\pi}(k_0,\vec{k})-\epsilon(k_0)\accentset{\circ}{\rho}_{\pi}(k_0,\vec{k})\right\} \nonumber\\
&=&-3\frac{m\Delta_m}{8f_\pi\bar{\sigma}}\int_0^\infty dk_0\int_0^\infty d|\vec{k}||\vec{k}|^2\left(F(|\vec{k}|,\Lambda)\right)^2\left\{\rho_{\pi}(k_0,\vec{k})-\accentset{\circ}{\rho}_{\pi}(k_0,\vec{k})\right\}\ .\label{Tadpole2}
\end{eqnarray}
\end{widetext}
In these equations, we have defined the two-point functions for $\sigma$ meson and pion in the vacuum:
\begin{eqnarray}
\accentset{\circ}{\tilde{G}}_\sigma(k_0,\vec{k}) &=& \frac{i}{k^2-\accentset{\circ}{m}_\sigma^2+i\epsilon} \ , \nonumber\\
\accentset{\circ}{\tilde{G}}_\pi(k_0,\vec{k}) &=& \frac{i}{k^2-\accentset{\circ}{m}_\pi^2+i\epsilon} \ .
\end{eqnarray}
We have utilized a fact that tadpole diagrams do not provide any imaginary parts and a property of $\rho_{\sigma(\pi)}(k_0,\vec{k}) = -\rho_{\sigma(\pi)}(-k_0,\vec{k})$. Combining Eqs.~(\ref{Tadpole1}) and~(\ref{Tadpole2}), we can get Eq.~(\ref{Hartree}).

Next, let us calculate the Fock-type one-loop diagram in Fig.~\ref{fig:DSelfEnergy} $(d)$. As explained in Sec.~\ref{sec:SymmetricMatter}, firstly we get the imaginary part, and secondly get the real part by using the subtracted dispersion relation. The imaginary part of retarded self-energy in Fig.~\ref{fig:DSelfEnergy} $(d)$ is obtained as (for the detail, see Ref.~\cite{Suenaga:2017deu})~\footnote{In this calculation, we redefine the zeroth-component of momenta of $\bar{D}$ meson and pion in such a way that the chemical potential does not appear in the indices, and this is true for calculations in Appendix.\ref{sec:OneLoopAS}}
\begin{widetext}
\begin{eqnarray}
{\rm Im}\tilde{\Sigma}_{{\rm Fig}. 4\, (d)}^R(q_0,\vec{q}) &=&\frac{1}{2}\left(\tilde{\Sigma}^>_{{\rm Fig}. 4\, (d)}(q_0,\vec{q})-\tilde{\Sigma}^<_{{\rm Fig}. 4\, (d)}(q_0,\vec{q})\right) \nonumber\\
&=&\frac{3}{2}\left(i\frac{m\Delta_m}{f_\pi}\right)^2\int\frac{d^4k}{(2\pi)^4}\left(F(|\vec{k}|,\Lambda)\right)^2 \nonumber\\
&&\times \Big\{\tilde{G}_{\pi}^>(q_0-k_0,\vec{q}-\vec{k}){\tilde{G}}^>_{\bar{D}}(k_0,\vec{k})-\tilde{G}_{\pi}^<(q_0-k_0,\vec{q}-\vec{k}){\tilde{G}}^<_{\bar{D}}(k_0,\vec{k})\Big\} \nonumber\\
&=&-\frac{3}{2}\left(\frac{m\Delta_m}{f_\pi}\right)^2\int\frac{d^4k}{(2\pi)^4}\left(F(|\vec{k}|,\Lambda)\right)^2 \nonumber\\
&&\times \Big\{\theta(q_0-k_0)\rho_\pi(q_0-k_0,\vec{q}-\vec{k})\theta(k_0){\rho}_{\bar{D}}(k_0,\vec{k})-\theta(k_0-q_0)\rho_\pi(q_0-k_0,\vec{q}-\vec{k})\theta(-k_0){\rho}_{\bar{D}}(k_0,\vec{k})\Big\} \nonumber\\
&=& -\frac{3}{2}\left(\frac{m\Delta_m}{f_\pi}\right)^2\int\frac{d^3k}{(2\pi)^3}\left(F(|\vec{k}|,\Lambda)\right)^2\frac{1}{2E_{\bar{D}}^k} \nonumber\\
&&\times \Big\{\theta(q_0-E_{\bar{D}}^k)\rho_\pi(q_0-E_{\bar{D}}^k,\vec{q}-\vec{k})+\theta(-q_0-E_{\bar{D}}^k)\rho_\pi(q_0+E_{\bar{D}}^k,\vec{q}-\vec{k})\Big\} \ ,\nonumber\\ \label{ImDem}
\end{eqnarray}
\end{widetext}
where $\tilde{\Sigma}_{{\rm Fig}. 4\, (d)}^>(q_0,\vec{q})$ and $\tilde{\Sigma}^<_{{\rm Fig}. 4\, (d)}(q_0,\vec{q})$ are the greater and lesser self-energy, respectively. ${\tilde{G}}^>_{\bar{D} (\pi)}(q_0,\vec{q})$ and ${\tilde{G}}^<_{\bar{D} (\pi)}(q_0,\vec{q})$ are the greater and lesser Green's function of $\bar{D}$ meson (pion), and these quantities are related to the spectral function ($\rho_{\bar{D}(\pi)}(q_0,\vec{q})$) as follows:
\begin{eqnarray}
{\tilde{G}}^>_{\bar{D} (\pi)}(q_0,\vec{q}) &=& \theta(q_0)\rho_{\bar{D}(\pi)}(q_0,\vec{q})\ ,\nonumber\\
\tilde{G}^<_{\bar{D}(\pi)}(q_0,\vec{q}) &=& -\theta(-q_0)\rho_{\bar{D}(\pi)}(q_0,\vec{q})\ .
\end{eqnarray}
In obtaining the last line in Eq.~(\ref{ImDem}), we have used
\begin{eqnarray}
\rho_{\bar{D}}(q_0,\vec{q}) = 2\pi\epsilon(q_0)\delta(q^2-M_{\bar{D}}^2)\ .
\end{eqnarray}
The final expression in Eq.~(\ref{ImDem}) is nothing but Eq.~(\ref{ImD0st}), and the real part is obtained by the subtracted dispersion relation in Eq.~(\ref{ReD0st}).

\section{Calculations of one-loop diagrams in Fig.~\ref{fig:DSelfEnergy} in asymmetric nuclear matter}
\label{sec:OneLoopAS}
In this appendix, we show self-energies of $\bar{D}_{0,u}^*$ and $\bar{D}_{0,d}^*$ mesons in asymmetric nuclear matter. The expression of the Hartree-type one-loop diagram in Fig.~\ref{fig:DSelfEnergy} ($b$) is unchanged from Eq.~(\ref{Tadpole1}). 
The result of diagram in Fig.~\ref{fig:DSelfEnergy} ($c$) can differ from Eq.~(\ref{Tadpole2}) since it includes pion loops. This is modified as 
\begin{widetext}
\begin{eqnarray}
\tilde{\Sigma}_{{\rm Fig}. 4\, (c)}  &=& - \frac{m\Delta_m}{2f_\pi\bar{\sigma}}\int\frac{d^4k}{(2\pi)^4}\left(F(|\vec{k}|,\Lambda)\right)^2\left\{ \tilde{G}_{\pi^0}(k_0,\vec{k})-\accentset{\circ}{\tilde{G}}_{\pi^0}(k_0,\vec{k})\right\}  \nonumber\\
&-&  \frac{m\Delta_m}{f_\pi\bar{\sigma}}\int\frac{d^4k}{(2\pi)^4}\left(F(|\vec{k}|,\Lambda)\right)^2\left\{ \tilde{G}_{\pi^+}(k_0,\vec{k})-\accentset{\circ}{\tilde{G}}_{\pi^+}(k_0,\vec{k})\right\}  \nonumber\\
&=& -\frac{m\Delta_m}{4f_\pi\bar{\sigma}}\int\frac{d^4k}{(2\pi)^4}\left(F(|\vec{k}|,\Lambda)\right)^2\left\{\epsilon(k_0)\rho_{\pi^0}(k_0,\vec{k})-\epsilon(k_0)\accentset{\circ}{\rho}_{\pi^0}(k_0,\vec{k})\right\} \nonumber\\
&-&\frac{m\Delta_m}{2f_\pi\bar{\sigma}}\int\frac{d^4k}{(2\pi)^4}\left(F(|\vec{k}|,\Lambda)\right)^2\left\{\epsilon(k_0)\rho_{\pi^+}(k_0,\vec{k})-\epsilon(k_0)\accentset{\circ}{\rho}_{\pi^+}(k_0,\vec{k})\right\} \ .
\end{eqnarray}
This is further simplified by making use of relations of $\rho_{\pi^0}(q_0)=-\rho_{\pi^0}(-q_0)$ and $\rho_{\pi^+}(q_0)=-\rho_{\pi^-}(-q_0)$:
\begin{eqnarray}
\tilde{\Sigma}_{{\rm Fig}. 4\, (c)}  &=&-\frac{m\Delta_m}{8f_\pi\bar{\sigma}}\int_0^\infty dk_0\int_0^\infty d|\vec{k}||\vec{k}|^2\left(F(|\vec{k}|,\Lambda)\right)^2 \nonumber\\
&&\times \left\{\rho_{\pi^0}(k_0,\vec{k})+\rho_{\pi^+}(k_0,\vec{k})+\rho_{\pi^-}(k_0,\vec{k}) -\accentset{\circ}{\rho}_{\pi^0}(k_0,\vec{k})-\accentset{\circ}{\rho}_{\pi^+}(k_0,\vec{k})-\accentset{\circ}{\rho}_{\pi^-}(k_0,\vec{k})\right\}\ , \label{TadPole2AS}
\end{eqnarray}
where we have employed $\accentset{\circ}{\rho}_{\pi^0}(k_0,\vec{k})=\accentset{\circ}{\rho}_{\pi^+}(k_0,\vec{k})=\accentset{\circ}{\rho}_{\pi^-}(k_0,\vec{k})$. In isospin symmetric limit, differences among $\pi^0$, $\pi^+$ and $\pi^-$ vanish and we find $m_{\pi^0}^2 = m^2_{\pi^+}=m_{\pi^-}^2$ and $\rho_{\pi^0}(k_0,\vec{k})=\rho_{\pi^+}(k_0,\vec{k})=\rho_{\pi^-}(k_0,\vec{k})$. Therefore, we can confirm Eq.~(\ref{TadPole2AS}) is identical to Eq.~(\ref{Tadpole2}) in the isospin symmetric limit.

Next, we compute the Fock-type diagram in Fig.~\ref{fig:DSelfEnergy} ($d$) in asymmetric nuclear matter. For $\bar{D}_{0,u}^*$ meson, this is of the form
\begin{eqnarray}
&& {\rm Im}\tilde{\Sigma}^{\bar{D}_{0,u}^*R}_{{\rm Fig}. 4\, (d)}(q_0,\vec{q}) \nonumber\\
&=&-\frac{1}{2}\left(\frac{m\Delta_m}{f_\pi}\right)^2\int\frac{d^4k}{(2\pi)^4}\left(F(|\vec{k}|,\Lambda)\right)^2 \left\{\tilde{G}_{\pi^0}^>(q_0-k_0,\vec{q}-\vec{k})\tilde{G}_{\bar{D}_u}^>(k_0,\vec{k})-\tilde{G}_{\pi^0}^<(q_0-k_0,\vec{q}-\vec{k})\tilde{G}_{\bar{D}_u}^<(k_0,\vec{k})\right\} \nonumber\\
&-& \frac{1}{2}\left(\sqrt{2}\frac{m\Delta_m}{f_\pi}\right)^2\int\frac{d^4k}{(2\pi)^4}\left(F(|\vec{k}|,\Lambda)\right)^2\left\{\tilde{G}_{\pi^+}^>(q_0-k_0,\vec{q}-\vec{k})\tilde{G}_{\bar{D}_d}^>(k_0,\vec{k})-\tilde{G}_{\pi^+}^<(q_0-k_0,\vec{q}-\vec{k})\tilde{G}_{\bar{D}_d}^<(k_0,\vec{k})\right\} \nonumber\\
&=&-\frac{1}{2}\left(\frac{m\Delta_m}{f_\pi}\right)^2\int\frac{d^3k}{(2\pi)^3}\left(F(|\vec{k}|,\Lambda)\right)^2 \frac{1}{2E_k^{\bar{D}}} \left\{\theta(q_0-E_k^{\bar{D}})\rho_{\pi^0}(q_0-E_k^{\bar{D}},\vec{q}-\vec{k})+\theta(-q_0-E_k^{\bar{D}})\rho_{\pi^0}(q_0+E_k^{\bar{D}},\vec{q}-\vec{k})\right\} \nonumber\\
&-&\left(\frac{m\Delta_m}{f_\pi}\right)^2\int\frac{d^3k}{(2\pi)^3}\left(F(|\vec{k}|,\Lambda)\right)^2 \frac{1}{2E_k^{\bar{D}}} \left\{\theta(q_0-E_k^{\bar{D}})\rho_{\pi^+}(q_0-E_k^{\bar{D}},\vec{q}-\vec{k})+\theta(-q_0-E_k^{\bar{D}})\rho_{\pi^+}(q_0+E_k^{\bar{D}},\vec{q}-\vec{k})\right\} \ .\nonumber\\
\end{eqnarray}
In the same way, for $\bar{D}_{0,d}^*$ meson, we find
\begin{eqnarray}
&& {\rm Im}\tilde{\Sigma}^{\bar{D}_{0,d}^*R}_{{\rm Fig}. 4\, (d)}(q_0,\vec{q}) \nonumber\\
&=&-\frac{1}{2}\left(\frac{m\Delta_m}{f_\pi}\right)^2\int\frac{d^4k}{(2\pi)^4}\left(F(|\vec{k}|,\Lambda)\right)^2 \left\{\tilde{G}_{\pi^0}^>(q_0-k_0,\vec{q}-\vec{k})\tilde{G}_{\bar{D}_d}^>(k_0,\vec{k})-\tilde{G}_{\pi^0}^<(q_0-k_0,\vec{q}-\vec{k})\tilde{G}_{\bar{D}_d}^<(k_0,\vec{k})\right\} \nonumber\\
&-& \frac{1}{2}\left(\sqrt{2}\frac{m\Delta_m}{f_\pi}\right)^2\int\frac{d^4k}{(2\pi)^4}\left(F(|\vec{k}|,\Lambda)\right)^2\left\{\tilde{G}_{\pi^-}^>(q_0-k_0,\vec{q}-\vec{k})\tilde{G}_{\bar{D}_u}^>(k_0,\vec{k})-\tilde{G}_{\pi^-}^<(q_0-k_0,\vec{q}-\vec{k})\tilde{G}_{\bar{D}_u}^<(k_0,\vec{k})\right\} \nonumber\\
&=&-\frac{1}{2}\left(\frac{m\Delta_m}{f_\pi}\right)^2\int\frac{d^3k}{(2\pi)^3}\left(F(|\vec{k}|,\Lambda)\right)^2 \frac{1}{2E_k^{\bar{D}}} \left\{\theta(q_0-E_k^{\bar{D}})\rho_{\pi^0}(q_0-E_k^{\bar{D}},\vec{q}-\vec{k})+\theta(-q_0-E_k^{\bar{D}})\rho_{\pi^0}(q_0+E_k^{\bar{D}},\vec{q}-\vec{k})\right\} \nonumber\\
&-&\left(\frac{m\Delta_m}{f_\pi}\right)^2\int\frac{d^3k}{(2\pi)^3}\left(F(|\vec{k}|,\Lambda)\right)^2 \frac{1}{2E_k^{\bar{D}}} \left\{\theta(q_0-E_k^{\bar{D}})\rho_{\pi^-}(q_0-E_k^{\bar{D}},\vec{q}-\vec{k})+\theta(-q_0-E_k^{\bar{D}})\rho_{\pi^-}(q_0+E_k^{\bar{D}},\vec{q}-\vec{k})\right\} \ .\nonumber\\
\end{eqnarray}
\end{widetext}
As is the case in the Hartree-type one-loop calculation, this result is reduced to the one in Eq.~(\ref{ImDem}) in the isospin symmetric limit. The real parts are obtained via the subtracted dispersion relation in Eq.~(\ref{ReD0st}).


\begin{thebibliography}{99}



\bibitem{Hatsuda:1994pi} 
See, e.g., T.~Hatsuda and T.~Kunihiro,
  Phys.\ Rept.\  {\bf 247}, 221 (1994)
  doi:10.1016/0370-1573(94)90022-1, as a review and references therein.
 
 

\bibitem{DeTar:2009ef} 
See, e.g., C.~DeTar and U.~M.~Heller,
  Eur.\ Phys.\ J.\ A {\bf 41}, 405 (2009)
  doi:10.1140/epja/i2009-10825-3, as a review and references therein.
    
    
\bibitem{Hayano:2008vn} 
See, e.g., R.~S.~Hayano and T.~Hatsuda,
  Rev.\ Mod.\ Phys.\  {\bf 82}, 2949 (2010), as a review and references therein.
  
  
\bibitem{Suzuki:2002ae} 
  K.~Suzuki {\it et al.},
  Phys.\ Rev.\ Lett.\  {\bf 92}, 072302 (2004)
  doi:10.1103/PhysRevLett.92.072302.



\bibitem{Suenaga:2014dia} 
  D.~Suenaga, B.~R.~He, Y.~L.~Ma and M.~Harada,
  Phys.\ Rev.\ C {\bf 89}, no. 6, 068201 (2014)
  doi:10.1103/PhysRevC.89.068201.

\bibitem{Suenaga:2014sga} 
  D.~Suenaga, B.~R.~He, Y.~L.~Ma and M.~Harada,
  Phys.\ Rev.\ D {\bf 91}, no. 3, 036001 (2015).

\bibitem{Suenaga:2015daa} 
  D.~Suenaga and M.~Harada,
  Phys.\ Rev.\ D {\bf 93}, no. 7, 076005 (2016)
  doi:10.1103/PhysRevD.93.076005.

   
\bibitem{Harada:2016uca} 
  M.~Harada, Y.~L.~Ma, D.~Suenaga and Y.~Takeda,
  Progress of Theoretical and Experimental Physics, Volume 2017,
  Issue 11, 1 November 2017, 113D01
  doi:10.1093/ptep/ptx140.
  
  
  
\bibitem{Suenaga:2017deu} 
  D.~Suenaga, S.~Yasui and M.~Harada,
  Phys.\ Rev.\ C {\bf 96}, no. 1, 015204 (2017)
  doi:10.1103/PhysRevC.96.015204.
  
 

\bibitem{HQSS}
For a review, see, e.g., Refs.~\cite{Neubert:1993mb,Manohar:2000dt} for the heavy quark physics and Ref.~\cite{Casalbuoni:1996pg} for applications of the heavy quark symmetry to the heavy hadron physics.

\bibitem{Neubert:1993mb}
 M.~Neubert,
 Phys.\ Rept.\  {\bf 245}, 259 (1994)
 [hep-ph/9306320].

\bibitem{Manohar:2000dt}
 A.V.~Manohar and M.B.~Wise,
 Camb.\ Monogr.\ Part.\ Phys.\ Nucl.\ Phys.\ Cosmol.\  {\bf 10}, 1 (2000).
 
  \bibitem{Casalbuoni:1996pg} 
	  R.~Casalbuoni, A.~Deandrea, N.~Di Bartolomeo, R.~Gatto,
	  F.~Feruglio and G.~Nardulli, 
	  Phys.\ Rept.\ {\bf 281}, 145 (1997).



\bibitem{Mishra:2003se}
  A.~Mishra, E.~L.~Bratkovskaya, J.~Schaffner-Bielich, S.~Schramm and H.~Stocker,
  Phys.\ Rev.\ C {\bf 69}, 015202 (2004).

    
\bibitem{Yasui:2009bz} 
  S.~Yasui and K.~Sudoh,
  Phys.\ Rev.\ D {\bf 80}, 034008 (2009)
  doi:10.1103/PhysRevD.80.034008.

\bibitem{Blaschke:2011yv} 
  D.~Blaschke, P.~Costa and Y.~L.~Kalinovsky,
  Phys.\ Rev.\ D {\bf 85}, 034005 (2012)
  doi:10.1103/PhysRevD.85.034005.
    
\bibitem{Yasui:2012rw} 
  S.~Yasui and K.~Sudoh,
  Phys.\ Rev.\ C {\bf 87}, no. 1, 015202 (2013)
  doi:10.1103/PhysRevC.87.015202.
  
\bibitem{Sasaki:2014asa} 
  C.~Sasaki,
  Phys.\ Rev.\ D {\bf 90}, no. 11, 114007 (2014)
  doi:10.1103/PhysRevD.90.114007.

  
\bibitem{Hayashigaki:2000es} 
  A.~Hayashigaki,
  Phys.\ Lett.\ B {\bf 487}, 96 (2000)
  doi:10.1016/S0370-2693(00)00760-7.
  
     
\bibitem{Hilger:2008jg} 
  T.~Hilger, R.~Thomas and B.~Kampfer,
  Phys.\ Rev.\ C {\bf 79}, 025202 (2009)
  doi:10.1103/PhysRevC.79.025202.

\bibitem{Azizi:2014bba} 
  K.~Azizi, N.~Er and H.~Sundu,
  Eur.\ Phys.\ J.\ C {\bf 74}, 3021 (2014)
  doi:10.1140/epjc/s10052-014-3021-1.
  
\bibitem{Wang:2015uya} 
  Z.~G.~Wang,
  Phys.\ Rev.\ C {\bf 92}, no. 6, 065205 (2015)
  doi:10.1103/PhysRevC.92.065205.
 
   
\bibitem{Suzuki:2015est} 
  K.~Suzuki, P.~Gubler and M.~Oka,
  Phys.\ Rev.\ C {\bf 93}, no. 4, 045209 (2016)
  doi:10.1103/PhysRevC.93.045209.
    
  
\bibitem{Lutz:2005vx} 
  M.~F.~M.~Lutz and C.~L.~Korpa,
  Phys.\ Lett.\ B {\bf 633}, 43 (2006)
  doi:10.1016/j.physletb.2005.11.046.


\bibitem{Tolos:2009nn} 
  L.~Tolos, C.~Garcia-Recio and J.~Nieves,
  Phys.\ Rev.\ C {\bf 80}, 065202 (2009).
  
\bibitem{Gamermann:2010zz} 
  D.~Gamermann, C.~Garcia-Recio, J.~Nieves, L.~L.~Salcedo and L.~Tolos,
  Phys.\ Rev.\ D {\bf 81}, 094016 (2010).
  

\bibitem{JimenezTejero:2011fc} 
  C.~E.~Jimenez-Tejero, A.~Ramos, L.~Tolos and I.~Vidana,
  Phys.\ Rev.\ C {\bf 84}, 015208 (2011)
  doi:10.1103/PhysRevC.84.015208.

  
  

\bibitem{GarciaRecio:2011xt} 
  C.~Garcia-Recio, J.~Nieves, L.~L.~Salcedo and L.~Tolos,
  Phys.\ Rev.\ C {\bf 85}, 025203 (2012).



\bibitem{Tsushima:1998ru} 
  K.~Tsushima, D.~H.~Lu, A.~W.~Thomas, K.~Saito and R.~H.~Landau,
  Phys.\ Rev.\ C {\bf 59}, 2824 (1999)
  doi:10.1103/PhysRevC.59.2824.

\bibitem{Mishra:2008cd} 
  A.~Mishra and A.~Mazumdar,
  Phys.\ Rev.\ C {\bf 79}, 024908 (2009)
  doi:10.1103/PhysRevC.79.024908.
  


     
\bibitem{Kumar:2010gb} 
  A.~Kumar and A.~Mishra,
  Phys.\ Rev.\ C {\bf 81}, 065204 (2010)
  doi:10.1103/PhysRevC.81.065204.
  
\bibitem{Kumar:2011ff} 
  A.~Kumar and A.~Mishra,
  Eur.\ Phys.\ J.\ A {\bf 47}, 164 (2011)
  doi:10.1140/epja/i2011-11164-6.
  
  
\bibitem{Chhabra:2016vhp} 
  R.~Chhabra and A.~Kumar,
  Eur.\ Phys.\ J.\ A {\bf 53}, no. 5, 105 (2017)
  doi:10.1140/epja/i2017-12285-6.
  
 

   
\bibitem{Hattori:2015hka} 
  K.~Hattori, K.~Itakura, S.~Ozaki and S.~Yasui,
  Phys.\ Rev.\ D {\bf 92}, no. 6, 065003 (2015)
  doi:10.1103/PhysRevD.92.065003.
  
                  
\bibitem{Hosaka:2016ypm} 
  A.~Hosaka, T.~Hyodo, K.~Sudoh, Y.~Yamaguchi and S.~Yasui,
  Prog.\ Part.\ Nucl.\ Phys.\  {\bf 96}, 88 (2017)
  doi:10.1016/j.ppnp.2017.04.003.
 



\bibitem{Nowak:1992um} 
  M.~A.~Nowak, M.~Rho and I.~Zahed,
  Phys.\ Rev.\ D {\bf 48}, 4370 (1993)
  doi:10.1103/PhysRevD.48.4370.
  
\bibitem{Bardeen:1993ae}
  W.~A.~Bardeen and C.~T.~Hill,
  Phys.\ Rev.\  D {\bf 49}, 409 (1994).
  

  
\bibitem{Detar:1988kn}
  C.~E.~Detar and T.~Kunihiro,
  Phys.\ Rev.\ D {\bf 39}, 2805 (1989).  

\bibitem{Nemoto:1998um} 
  Y.~Nemoto, D.~Jido, M.~Oka and A.~Hosaka,
  Phys.\ Rev.\ D {\bf 57}, 4124 (1998).

\bibitem{Jido:1998av} 
  D.~Jido, Y.~Nemoto, M.~Oka and A.~Hosaka,
  Nucl.\ Phys.\ A {\bf 671}, 471 (2000).

\bibitem{Jido:2001nt} 
  D.~Jido, M.~Oka and A.~Hosaka,
  Prog.\ Theor.\ Phys.\  {\bf 106}, 873 (2001).

\bibitem{Jido:1999hd} 
  D.~Jido, T.~Hatsuda and T.~Kunihiro,
  Phys.\ Rev.\ Lett.\  {\bf 84}, 3252 (2000)
  doi:10.1103/PhysRevLett.84.3252.



\bibitem{Motohiro:2015taa} 
  Y.~Motohiro, Y.~Kim and M.~Harada,
  Phys.\ Rev.\ C {\bf 92}, no. 2, 025201 (2015)
  Erratum: [Phys.\ Rev.\ C {\bf 95}, no. 5, 059903 (2017)]
  doi:10.1103/PhysRevC.92.025201, 10.1103/PhysRevC.95.059903.



\bibitem{Suenaga:2017wbb} 
  D.~Suenaga,
  Phys.\ Rev.\ C {\bf 97}, no. 4, 045203 (2018)
  doi:10.1103/PhysRevC.97.045203.

\bibitem{Harada:2003jx} 
  M.~Harada and K.~Yamawaki,
  Phys.\ Rept.\  {\bf 381}, 1 (2003)
  doi:10.1016/S0370-1573(03)00139-X.


\bibitem{Kapusta:2006pm} 
  J.~I.~Kapusta and C.~Gale, {\it Finite-Temperature Field Theory: Principles and Applications}, Cambridge Monographs on Mathematical Physics (Cambridge
University Press, Cambridge, 2006).
     

\bibitem{TFT}
See, e.g., M. Le Bellac, {\it Thermal Field Theory},
Cambridge Monographs on Mathematical Physics (Cambridge
University Press, Cambridge, 2000).



  
   
\bibitem{Chiku:1997va} 
  S.~Chiku and T.~Hatsuda,
  Phys.\ Rev.\ D {\bf 57}, 6 (1998)
  doi:10.1103/PhysRevD.57.6.

\bibitem{Hidaka:2002xv} 
  Y.~Hidaka, O.~Morimatsu and T.~Nishikawa,
  Phys.\ Rev.\ D {\bf 67}, 056004 (2003)
  doi:10.1103/PhysRevD.67.056004.

  

\bibitem{Yamagata-Sekihara:2015ebw} 
  J.~Yamagata-Sekihara, C.~Garcia-Recio, J.~Nieves, L.~L.~Salcedo and L.~Tolos,
  Phys.\ Lett.\ B {\bf 754}, 26 (2016)
  doi:10.1016/j.physletb.2016.01.003.


\bibitem{Shyam:2016bzq} 
  R.~Shyam and K.~Tsushima,
  Phys.\ Rev.\ D {\bf 94}, no. 7, 074041 (2016)
  doi:10.1103/PhysRevD.94.074041.
     
\end{thebibliography}
\end{document}